**Evaluating Intraspecific Variation And Interspecific Diversity:**
**Comparing Humans and Fish Species**
Bradly John Alicea

# ABSTRACT

The idea of what constitutes an animal species has been a topic of longstanding interest in the biological sciences. Therefore, the analysis of eight molecular datasets involving human and teleost examples along with morphological samples from several groups of Neotropical electric fish (Order: Gymnotiformes) were used in this thesis to test the dynamics of both intraspecific variation and interspecific diversity.

Three questions concerning both intraspecific and interspecific subdivision were addressed in this thesis. Meaningful intraspecific divisions were assessed by testing whether or not modern human variation can be subdivided into geographically distinct groups as was shown among two fish species. This was further tested by observing trends in a worldwide distribution of three craniometric traits within *Homo sapiens*. In terms of investigating molecular interspecific diversity among humans, two experimental exercises were performed. A cladistic exchange experiment tested for the extent of discontinuity and interbreeding between *H. sapiens* and neanderthal populations. As part of the same question, another experimental exercise tested the amount of molecular variance resulting from simulations which treated neanderthals as being either a local population of modern humans or as a distinct subspecies. Finally, comparisons of hominid populations over time with fish species helped to define what constitutes taxonomically relevant differences between morphological populations as expressed among both trait size ranges and through growth patterns that begin during ontogeny.

Compared to the subdivision found within selected teleost species, *H. sapiens* molecular data exhibited little variation and discontinuity between geographical regions. This conclusion was based on the results of phylogenetic analyses and molecular variance measures. Results of the two experimental exercises concluded that neanderthals exhibit taxonomic distance from modern *H. sapiens*. However, this distance was not so great as to exclude the possibility of interbreeding between the two subspecific groups. Finally, a series of characters were analyzed among species of Neotropical electric fish. These analyses were compared with hominid examples to determine what constituted taxonomically relevant differences between populations as expressed among specific morphometric traits that develop during the juvenile phase. Such examples were ultimately useful for purposes of objectively defining subdivisions among populations of *H. sapiens*.

# INTRODUCTION

This thesis sought to establish what constitutes both intraspecific variation and interspecific diversity. Examples from two disparate groups of animals--hominids and teleosts--were useful in a comparative approach which relied upon taxonomy to interpret subdivisions found within a given population.

Three specific questions were taken into consideration. The first question regarded intraspecific variation among four modern human and four teleost datasets. These datasets were useful for building a model that determined the nature of intraspecific diversity. This model along with a worldwide sample of three craniometric traits from Howells (1973) was specifically used to examine the existence of intraspecific subdivisions within *Homo sapiens*. These findings also provided a broad framework for determining the nature of intraspecific discontinuities using molecular data.

The existence of well-supported phylogenetic groups and between group variance was also used as the defining features of molecular interspecific diversity. Using only human data, a second question was asked in the form of two experimental exercises which tested the taxonomic relationship between modern human and neanderthal molecular data. For purposes of this thesis, it was assumed that if neanderthal constituted a distinct subspecies then they also had some capacity to interbreed with modern humans. This thesis attempted to statistically test the taxonomic status of neanderthal in relation to modern humans through two experimental exercises.

The first exercise involved determining the approximate number of mutational steps between neanderthals and modern humans. This allowed for the extent of reproductive isolation between these two distinct populations to be inferred. Limited interbreeding between neanderthal and modern human was tested for by moving single groups of modern human specimens both into a group formed by the neanderthal specimens and in between all other human groups. Total tree lengths for each permutation was recorded and produced a qualitative probability for interbreeding. A second exercise involved interpreting molecular variance measurements. This detection of taxonomic variance was then used to assess the taxonomic status of a potential neanderthal subspecies.

The third area of inquiry involved detecting taxonomically relevant differences among meristic and metric morphological characters in an interspecific context. Meristic characters are those which are discrete and can be counted, while metric characters are those that can be measured. In order to determine what distinguishes between taxonomic groups in a morphological context, three issues were addressed in an attempt to understand the full complexity of interspecific relationships. First, the limits of a species over time were determined through looking at selected traits from specimens across the range of *Homo erectus*. Next, the size range for two meristic characters among species groupings of *Gymnotus* were compared interspecifically for any potential discontinuity between species groups. Finally, the nature of distinguishing between taxonomic groups for morphometric traits among interspecific samples was explored using heterochrony as a theoretical framework. Interspecific comparisons were made using data from Neotropical electric fish (Order: Gymnotiformes) species, while applications of heterochrony to hominid populations was also discussed.

In general, morphological traits in both humans and fish grow and develop in similar ways between populations over the course of evolution. For example, some body parts are traditionally conserved over long periods of geologic time. However, other selected traits can vary widely in terms of size and shape across space and time. Meristic characters are less relevant to direct comparisons between humans and fish, for only differences in quantity act to differentiate between taxonomic groups. These traits have been used to subdivide gymnotiform diversity, but were not a primary form of evidence[1].

Even though there are innumerable differences between fish and humans, the conclusions reached by this thesis relied upon a number of generalizations. For example, Avise (2000) has argued that *H. sapiens* follows a pattern similar to that of *Anguilla rostrata* (American Eel). Each of these species were found to exhibit a high degree of gene flow and relatively small effective population sizes. As a result, a substantial lack of interpopulational geographical structuring has occured in both species. This has also resulted in a fewer number of significant polymorphic differences within *H. sapiens*, with newly arisen variants being more localized in nature (Avise *et al.* 1987). By using the same set of analyses on populations exhibiting various geographic distributions, strict comparisons can be made between vastly dissimilar taxa.

**Basis for Approach to Research Questions**

The tendency to either lump or split biodiversity into discrete taxonomic categories has been around since Linneaus proposed that both humans and great Apes be placed in the genus *Homo* (Cain 1954). Weidenreich and Broom took lumping to an extreme in the mid-20th century by proposing that all *Homo* specimens were part of a single evolving species (Tattersall 2000). Campbell (1963) supported the case for splitting by suggesting that *Homo* can be subdivided into a number of interspecific groups based on fossil evidence. Yet Strait *et al.* (1997) found that when the genus *Homo* is split up into more than four species, evolutionary relationships within *Homo* become obscured.

In the 1960's, Goodman used cladistic methods to measure immunological distances (Tattersall 2000). Based on a lack of differences, he suggested that both *Homo* and all great ape species belong in the family Hominidae. Tattersall (2000) has concluded that in general the preference to lump or split is not based on the type of available evidence. Thus, the following section introduces the nature of biodiversity and how it has been subdivided into formal biological units.

Three principles were used to answer the questions central to this thesis: the definition of taxonomic categories and how polymorphisms define those groups, the difference between biospecies and chronospecies, and the difference between panmictic and discontinuous populations. These factors all play an important role in defining both intraspecific and interspecific subdivision.

**Polymorphism and Types of Species**

Mayr (1954) has pointed out that a species can be highly polymorphic in nature. Polymorphic species differ from polytypic species in that the latter consist of clearly differentiated subspecies (Mayr 1963). Stadler *et al.* (2000) point out that differential phenotypes are products of mutation and patterns of organismal development. Therefore, the occurrence of several discontinuous phenotypes may be an expression of intraspecific variation (Yablokov 1986). Cain (1954) points out that modern taxonomy as proposed by Linneaus merely formalized traditional European folk classifications. While these classifications were sufficient in defining obvious morphological differences common to Europe, they were poor in identifying cryptic species or subtle differences in variation.

**Chronospecies and Biospecies**

There are two types of species based on whether or not specimens are sampled from different points in time. Since species are both spatially and temporally bounded entities, there is discreteness in any given biota (Lieberman and Vrba 1995). A biospecies can be defined as extant taxa, and can be characterized by polymorphism and differentiation only in a geographic context (Mayr 1954). This provides a view of diversity from a single point in time. The gymnotiform comparative material are characteristic of this species type. Variation in a chronospecies is characterized by both changes in space and time (Simpson 1961). The boundary of chronospecies are defined by both stratigraphy and geography. For example, the neanderthal molecular samples analyzed in this thesis were members of a chronospecies. It is also important to note that in the case of both chronospecies and biospecies, the production of a lineage does not automatically transform local populations into new species (Wiley 1981).

Willi Hennig's (1966) original view of chronospecies was that every modern species and their hypothetical ancestors represent different species. However, this does not take into account intraspecific changes over time (Nelson and Platnick 1981). Huelsenbeck (1991) claims that this is the case only when there are very few extant taxa in the analysis. Indeed, when fossil species are juxtaposed with biospecies in a phylogenetic context, methods should be developed to convey information about a taxon's temporal properties (De Queiroz and Donaghue 1988). One solution to this problem is that of stratocladistics (Fisher 1994). In this method, geological information is assigned a phylogenetic character state and added to the analysis (Beatty 1994). While this approach seems to be informative, it does not completely define the boundaries of a chronospecies.

**Panmictic Versus Discontinuous Species**

A single species can be either panmictic or discontinuous. When a given species is panmictic, it means that a group of organisms are entirely reproductively connected across their geographic range. All peripheral populations are randomly and fully connected to the core population in a series of weblike connections with a core population. Meanwhile, discontinuous population distributions exhibit a lack of web-like relationships to some extent over a portion of its range (Avise 2000). When there are barriers to the reproductive interconnectedness of two or more populations over time, subdivision within and between taxonomic groups is the result.

Intraspecific and interspecific analyses differ in that intraspecific populations do not behave like whole species. Wright (1956) originally saw species as being broken up into one or more local populations called colonies. These quasi-isolated local populations are referred to as demes if they are reproductively active. If affinities between these groups are lacking, it can be used to support the existence of either subspecific or species-level groups. Wright's (1956) shifting balance theory actually considers demic differentiation over short periods of time to be the opposite of speciation. Thus even though they may act as important sources of diversity, demes are never totally isolated from other interspecific populations.

**Cultural Constraints on Intraspecific Variation**

Gene flow acts to shape intraspecific variation, and can be estimated by measuring the variance within populations and between groups. When a species exhibits discontinuity, a larger number of segregated sites between groups than within groups can be expected (Slatkin 1987). Excoffier *et al.* (1992) conducted a hierarchical analysis of variance (AMOVA), and demonstrated that a majority of variability within *H. sapiens* was present among small demes. This translates into a general lack of between group variation across the geographic range of modern humans.

The analysis of human populations is unique in that ethnic groups can be used to represent demic populations. Occasionally, human populations can be classified as what Excoffier *et al.* (1992) call "contrived regional groups". These groups exhibit large amounts of internal divergence, and are primarily defined by geographic, linguistic, or religious boundaries (Stoneking et al. 1990). Thus, ethnic groups can be marked with varying degrees of molecular diversity which may or may not be analogous to what is found within single fish species. The Ngobe of Panama (Kolman *et al.* 1996) exhibit a particularly low degree of diversity. By contrast, the !Kung and Pygmy are illustrative of extensive geographic structuring and high diversity within a single ethnic group (Vigilant *et al.* 1989).

One of the most important cultural factors is linguistic differentiation. In terms of specific case studies, Greenberg *et al.* (1986) found that native New World populations characterized by three waves of initial colonization relate to the three main linguistic groups found in the Americas today. While linguistic factors are no doubt important, they can occasionally conflict with established molecular findings.

This is exemplified by two tree diagrams presented by Penny *et al.* (1993). One topology was based solely on linguistic data while the other was exclusively a molecular tree, which resulted in two substantially different topologies. For example, Penny *et al.* (1993) found that molecular sequences from Iranians and Berbers belonged to the same group. When linguistic evolution was taken into account, however, the Berbers and Iranians ended up in separate groups. Penny *et al.* (1993) also found much the same result when comparing the linguistic and molecular affinities between Asian Indians and Dravidians.

**Footnotes:**

[1]- Since trait size and shape is naturally variable within a single species due to stochastic variables, change in these traits must show some discernible trend. In addition, the interpretation of trait size and shape through theoretical concepts such as heterochrony and homoplasy resulted in a generalized conclusion applicable to broad comparisons.

# METHODOLOGICAL BACKGROUND

This section provides a background to the analyses and data used in this thesis. The theory of systematization and procedures behind the AMOVA method were introduced first. This information was either applied directly to the data or used as background to understand morphological and molecular analyses. Standard definitions of intraspecific subdivision and an introduction to neanderthal populations yields to a description of gymnotiform diversity. In addition, the concept of heterochrony was also introduced as a framework for analyzing the interspecific growth rates of diagnostic characters. Finally, an introduction to and review of four major species concepts were covered.

**Methods of Systematization**

The interrelationships between intraspecific and interspecific populations were interpreted using the methods shown in Table 1.

**Table 1:** A comparison of two main types of systematization (Schuh 2000).

|  | **Numerical Taxonomy** | **Cladistics** |
|---|---|---|
| **Data Type** | discrete character distances | discrete character states |
| **Grouping Method** | overall similarity | unique similarity |
| **Hierarchical Level Determined By** | amount of difference | sharing of unique attributes |

Griffiths (1974) and Simpson (1961) have recognized that classification and systematization are two distinct types of ordering activities which can be uncoupled. Furthermore, distinguishing between systematization, classification, and their products is what O'Hara (1993) presents as central to defining distinct taxonomic groups.

**Cladistics and its Terminology**

Phylogenetic inference can best be defined as a hypothetical representation of evolutionary events and biological units as they may have been interrelated in the past (Sober 1988). Hennig (1966) believed that there should be a one-to-one correspondence between systematization and taxonomy where derived taxa are always nested in clades rooted by their hypothetical ancestors. However, Scott-Ram (1990) has observed that cladograms and classifications do not have similar logical structures. Cladistics treats each trait as a single character value in a series ranging from an ancestral to a derived state. The ancestral nature of a trait can be determined through comparative anatomy or including an outgroup into the analysis. Once the ancestral nature of a trait is determined, it can be said to be polarized (Kitching *et al.* 2000). Polarization makes it much easier to determine the evolutionary significance of a given trait.

In cladistic terminology, a single character can be evolutionarily significant in three main ways. A synapomorphic character is a shared derived character which serves to unite two or more taxa into a monophyletic group (Schuh 2000). Meanwhile, an autapomorphy is a derived character restricted to a single taxon. All characters in a cladistic analysis should be homologous. There are two tests which determine whether or not a trait can be classified as such. The first test requires characters to exhibit a degree of similarity among all organisms in a given analysis through their presence, absence, or transformation. Any trait, from a single bone to a nucleotide base, can be regarded as a suitable homologous structure (Wagner 1994). Secondly, a character must not have arrived at their current state by means of convergent evolution. If characters do not pass this second test, then they are said to be homoplastic.

Homoplastic characters have generally been considered to be uninformative for inferring evolutionary relationships. Nevertheless, homoplasy is an emerging issue in hominid evolution. Several investigators have drawn interesting conclusions by considering traditionally homoplastic characters as being informative. For example, Collard (in Wood 1999) found that masticatory characters traditionally thought to be homoplastic are actually comparable with other regions of the cranium. Lahr (1996) has even speculated that homoplasy exhibited among taxa may point to limited interbreeding in an interspecific context.

Finally, phylogeography is a means by which intraspecific phylogenies are interpreted through superimposition over local geography. This method was used to determine whether or not the expected mean number of site differences within a intraspecific population was panmictic, or informative across its entire geographical range (Templeton 1989). Used in tandem, phylogenies and geographic maps shouldyield breaks between haplotypes, as groups of individuals in the population differed from other such clades by many mutational steps. These populations would then occupy either separate geographical regions within the range of a species or were geographically concomitant.

**Numerical Taxonomy**

Numerical taxonomic methods use techniques that measure the degree to which groups are similar and different which is mediated by assuming constant rates of change (Felsenstein 1983). Using the Arlequin software package (Schneider *et al.* 2000), one can request the sample and group parameters of a given molecular dataset. Statistics are then produced which characterize the hierarchical structure of particular species and local populations in terms of reproductive cohesion, reproductive isolation, and the relative distance between subgroups (Schneider *et al.* 1999). When an analysis is performed, raw data representing molecular polymorphisms is transformed to produce distance measures among and between individuals, population groups, or even species. In a haploid genetic system, analyses of inter-haplotypic distances yielded *n* individuals from *i* populations that form the basis of a *n*-squared matrix. Individual cells in a matrix may represent higher levels of analysis such as individuals grouped by geography, ecosystem, or language group (Sneath and Sokal 1973).

**AMOVA and the Measure of Molecular Variance**

One method for determining population architecture is by performing an AMOVA. AMOVA, or Analysis of Molecular Variance, is best described as assigning a numerical value to the distance between polymorphisms that distinguish members of a given population. AMOVA provides a measure of the variance in gene frequencies while also taking into account the number of mutations between molecular haplotypes (Schneider *et al.* 2000). AMOVAs also produce a distance both between and within populations which can be expressed as a percentage of polymorphism within the entire dataset. A hierarchical analyses of molecular variance divides the total amount of variance in the dataset into covariance components. These components differ at the within populations, between populations, and between group levels of analysis (Excoffier *et al.* 1992). Each component is defined by a single parameter, and are defined by the coefficients one, two, and three respectively. Similar to other forms of multivariate analysis, the percentage of variance at each level represents the total amount of diversity in terms of the entire dataset that can be explained by focusing exclusively on a single covariance component at a single level (Excoffier *et al.* 1992). Each covariance component essentially represents a squared Euclidean distance between the units of comparison at each hierarchical level.

Summing every level of analysis in the dataset should yield a percentage of 100. While no negative components occurred in these analyses, their presence can be interpreted as a lack of structuring at that level. The concept of hierarchical partitions of variance is important in defining taxonomic discontinuity, because AMOVA treats geographically based samples and groups as constituting three distinct levels. Thus, a high degree of variance between groups or regions may have been more informative of potential subdivision than high variance levels within populations. Overall, the most important question when considering what AMOVA distances represent is the magnitude of each component relative to the other two levels of analysis.

**Background for Analyses of Intraspecific Variation**

Modern human intraspecific groupings are not clearly identified using either morphological or molecular evidence. Excoffier *et al.* (1992) discovered that 80.68 % of all haplotypic diversity within the species *H. sapiens* was to be found within each population subdivision, but only 15.73% of that diversity separated regional populations. Furthermore, the amount of difference between populations between populations from different continents was only 3.59 % (Excoffier *et al.* 1992). Overall, the evidence has consistently shown that because molecular variation within *H. sapiens* is large within populations as opposed to between groups, there is a lack of geographically-based discontinuity in the modern human species.

Shea *et al.* (1993) investigated craniometric variation within the chimpanzee species *Pan troglodytes* in order to better understand the nature of these morphological traits within *H. sapiens*. In this case, intraspecific designations were found to be useful as a heuristic tool rather than an actual category. More specifically, distinct intraspecific groups were supported by either a single trait or limited sets of characteristics. Such arbitrary criterion stands in contrast to Mayr (1963), who required that formal intraspecific categories truly reflect the breadth and patterning of potential variance in a species-level grouping. Ultimately, the application of intraspecific groups to analyses of non-human primate populations provided little more than superficial partitioning of a single species.

Molecular information also helps to elucidate relationships both within and between human groups. In this thesis, demographic patterns and genetic drift were deduced from specific markers in the mitochondrial genome known as haplotypes. The standard definition of a haplotype is a particular combination of alleles found together on a single haploid chromosome (Ward 1991). In theory and following Avise (2000), more haplotypes should be present in populations that have either been in sustained reproductive contact with a central core population or experienced significant population growth. By contrast, populations peripheral to that core or that have suffered a demographic collapse should exhibit fewer haplotypes and linkages as they are products of genetic drift (Avise *et al.* 1987).

**Description and Introduction to Neanderthal**

As a chronospecies, neanderthals lived from 130,000 to 35,000 years ago. The type specimen originates from the Neander Valley in Germany, and was found in 1847. neanderthal populations ranged over most of Europe, North Africa, and Western Asia. In terms of morphology, neanderthals are generally characterized by their large brow ridges, long skulls, large teeth, and face which is not placed underneath the brain (Trinkaus and Howells 1979). Neanderthals also exhibited a more robust body size on average than modern humans. Traditionally, grade-based definitions of  neanderthal have involved descriptions intermingling primitive and derived characters. According to Gowlett (1993), these characteristics evolved over time so that potentially diagnostic characteristics appeared closer to the time of neanderthal's demise.

Traditionally, there have been two competing perspectives on the relationship between neanderthals and modern humans. One hypothesis maintains that neanderthals and humans are extremes in the variation of a single species, separated only by time and geographic isolation due to shifts in the Pleistocene climate. A second explanation contends that neanderthals are a distinct subspecies within *H. sapiens*, and that a distinct taxonomic unit named *H. sapiens neanderthalensis* existed in reproductive isolation from modern humans during much of the late Pleistocene.

Whether neanderthal is considered a local population of *H. sapiens* or a distinct subspecies, there is no question that neanderthals exhibited culture. For example, a neanderthal burial from Shanidar Cave dated to 60,000 years ago was found in association with flower pollen. The general interpretation of this finding is that neanderthals exhibited some capability for ritual and symbolism (Trinkaus and Howells 1979). While neanderthal and modern humans both exhibited what could be called cultural behavior, there are distinct contrasts between the two groups based on archaeological evidence. Modern *H. sapiens* populations living in Europe and western Asia during the upper Pleistocene specialized in a stone blade tool technology rarely associated with neanderthal populations from roughly the same time period (Gowlett 1993). In addition, while modern humans are closely associated with elaborate cave art, no such examples have been associated with neanderthal.

**Description of Neotropical Electric Fish Diversity**

*Gymnotus and Sternopygus* are both gymnotiform genera, and represented the teleost interspecific morphological data used in this thesis. Variation among these organisms was marked in a number of ways. Meristic characters such as the number of vertebrae or rays were scored. Body measurements were also compared with the range for all other specimens in the analysis. Both metric and meristic data were used to differentiate between species.

The divergence of key morphological features that ultimately differentiate between species can be explained by the concept of heterochrony. According to this theory, the growth trend for a diagnostic character among individual members of a certain species can either begin earlier or later during a phase early in the life cycle called ontogeny. Physical growth then continues from the embryonic stage at either an accelerated or decelerated pace, all of which is relative to individuals in other species or populations (Buss 1987). This affects the lifelong growth trend, which in turn tends to produce a linear function. Such a trend is predictive to the extent that certain plastic traits will vary in terms of size range among morphologically different groups. Thus, heterochrony involves changes in the relative rates of timing in the development of characters interspecifically. Heterochrony also involves an uncoupling of size, shape and time variables. More specifically, ontogenetic changes in the size and shape of a character over time produced the phenomenon of heterochrony (McKinney and McNamara 1991).

When a specimen grew "much faster", its growth trajectory was noticeably different from all the other species on a bivariate graph. For example, mouth width growth began latest in *Gymnotus varzea*, but was accelerated so that over the lifespan of a specimen, this region grew faster than it does in *Gymnotus maculosus, Gymnotus pantherinus,* or *Gymnotus inaequilibiatus.*

## Major Species Concepts

Endler and Otte (1989) observed that species are tools that are fashioned for characterizing organic diversity. According to Hennig (1966), species groupings which take into account all intraspecific variation should be a prerequisite to taxonomic analysis. In a more ephemeral sense, O'Hara (1993) proposes that species are complex and historical entities.

Wheeler and Meier (2000) suggest two goals that should be central to the species concepts listed in Table 2. The first is to identify groups with a retrievable common history and which may not be divided into less inclusive units for which the same is true should be discovered. The second goal is to find the least inclusive units accorded formal recognition consistent with the goals of communicating and predicting the distribution of characters among organisms.

**Table 2:** Four major species concepts and their characteristics.

| Species Concept | Definition | Consequence |
|---|---|---|
| BSC (Biological Species Concept) | Isolated breeding populations | Isolated and discrete units in space |
| Phylogenetic | Least inclusive diagnosable lineage | Species diagnosed by "fixation" of a character state |
| Evolutionary | Independent ancestor-descendant lineage | Fixed and discrete units in time |
| Ecological | One species is representative of a single niche | Feeding structures become more informative, niches determine species |

The most popular species concept is that of Ernst Mayr. His seminal Biological Species Concept (BSC) looks at interbreeding natural populations and defines a species as any group that is reproductively isolated from other such groups. Mayr's definition is also primarily concerned with the spatial dimension. Wiley (1981) points out that exactly where a new species appears is causally related to a geographic disjunction which prevents interdemic migration. Eckhardt (2000) takes a similar view by defining biological species as a multicentered network that represents evolution with gene flow. Cain (1954) notes that when reproductive isolation cannot be observed, taxonomists must look at morphological or molecular differences between populations. For this reason, all groups which overlap must be assigned at least a subspecific designation.

According to Templeton (1981), isolation and cohesion are two sides of the same coin. Species defined as cohesive must possess and maintain distinct molecular lineages. Genetic exchangeability and mechanisms promoting genetic identity also defines the limits of a species through gene flow (Fisher 1930). Selective fixation is also a prime mover in cohesion, which means that natural selection promotes genetic identity by favoring the fixation of a genetic variant (Templeton 1981). This is one of the reasons why the subdivision of populations into localized demes does not automatically lead to the formation of new species.

Cracraft (1983) defines phylogenetic species as the smallest diagnosable cluster of individual organisms within which there is a parental pattern of ancestry and descent. Populational characteristics such as reproductive cohesion or isolation are never criterion in and of themselves for delimiting species. Instead, apomorphies and autapomorphies become most important (Wheeler and Meier 2000). The phylogenetic species concept is often used among Neotropical electric fish, for differentiating between species is mainly an act of differentiating between relative degrees of character fixation (Albert *et al.* 1999b). Yet in the case of molecular data,  Avise and Ball (1990) claim that there are large differences in nucleotide diversity values among different loci and for microsatellite diversity within a single species. At this level of analysis, the smallest phylogenetic species is much smaller than the largest interbreeding group.

The evolutionary species concept requires that an species entity maintains identity through space and over time. Simpson (1953) first developed this concept to deal with the temporal dimension of organismal distribution using exclusively fossil data as support. Wheeler and Meier (2000) define an evolutionary species as including the largest system that exists among individuals capable of producing offspring taking both space and time into account. This species concept has found its niche among paleontologists (Kimbel and Martin 1993), but provides little guidance as to which traits are most important in defining species (Templeton 1989).

Finally, the formal definition of the Ecological Species Concept as given by Wiley (1978) states that species units are unified not only by gene flow, but also through developmental, genetic, and other ecologically-specific causal mechanisms. Avise (2000) supports the viewpoint that low effective breeding group sizes in large modern populations may be the result of historical demographic fluctuations caused by climate change. Harpending *et al.* (1998) argued that modern humans exhibited a small effective population size of around 10,000 individuals throughout the harsh climatic conditions of the Pleistocene. Strumbauer *et al.* (2001) also focused on the African Pleistocene when they concluded that water levels are an important factor in the formation of new species in the Great Lakes of Africa among its populations of cichlid fishes. Regardless of causal mechanism, these species concepts were all found to influence interpretations of trait and geographic distributions presented by various datasets

# MATERIALS AND METHODS

All datasets analyzed here were chosen on the basis of their potential to express several trends in human and teleost evolution. The Johnson (2001), Kolman *et al.* (1996), and Ward *et al.* (1991) molecular samples were chosen to simulate the nature of local intraspecific populations. By contrast, the Vigilant *et al.* (1991) datasets were chosen to provide a species-wide look at human molecular diversity. To test for any potential intraspecific discontinuity in the human dataset, the two fish species originally sampled by Brunner *et al.* (2001) and Waters and Burridge (1999) were chosen.

The neanderthal samples were chosen to test for the existence of an interspecific relationship between themselves and modern humans when tested against molecular data from fish. The gymnotiform specimens provided a teleost interspecific morphological analogue for discussing ways of distinguishing between taxonomic groups and the process of heterochrony within the genus *Homo*.

## Description of Phylogenetic Analytical Methods

Bootstrap analyses of heuristic searches were also implemented in PAUP 3.1.1 (Swofford 1994) to determine most parsimonious interpretations of the molecular datasets presented. A heuristic search evaluates all possible minimum-length cladograms but is not guaranteed to find the most parsimonious solution (Kitching *et al.* 2000). A heuristic search makes several general assumptions about the matrix being searched. For example, the algorithm attempts to find the shortest overall topology, but only knows what this should look like through following several general conditions. Thus, the program may reach a local as opposed to a global minimum.

Bootstrap analyses of heuristic searches randomly delete some characters and reweigh the remaining characters over a series of iterative runs so that they ultimately form a single pseudoreplicate tree. After a specified number of replicates are built, a majority consensus of these trees are used to find the most parsimonious cladograms and assigns values to each recognizable clade. If a clade was supported a 95 % degree of support, then such a clade appeared among 95 % of all pseudoreplicates built (Kitching *et al.* 2000). These scores were used as the primary criterion for evaluating the robustness of discontinuity within a given dataset. All bootstrap analyses were based on parsimony and a 50% majority-rule consensus, and produced 100 replicates for every heuristic search performed. The heuristic search of all data matrices presented in this thesis utilized branch swapping and the tree-bisection-reconnection (TBR) algorithm.

**Descriptions of the Molecular Datasets**

The molecular variance components derived from datasets in this thesis were based on specific transition to transversion ratios. A transition to transversion ratio of 15 to 1 was used for all human datasets (Nei and Tamura 1993), while the fish datasets were weighted at a ratio of 4 to 1 (Alves-Gomes *et al.* 1995).

Ward *et al.* (1991) looked at the mitochondrial diversity of the Nuu-Chah-Nulth, which is a Native American tribe in British Columbia. This deme was considered to be fairly reproductively isolated, and thus is a good candidate for examining within-group molecular diversity. The Nuu-Chah-Nulth population itself consists of 14 bands which have lived in the same region for 4000 years. Kolman *et al.* (1996) sampled at the regional scale and looked at mitochondrial diversity among two distinct ethnic populations in Mongolia. Kolman *et al.* (1996) sampled individuals belonging to the Khalkha and Dariganga ethnic groups. The Khalka represent 80 % of the Mongolian nation, while the Dariganga make up only 1.5 % of the country's population (Kolman *et al.* 1996). Although these two populations represent several demic populations within a much larger local population, it is estimated that the Dariganga only emerged as a separate population around 300 years ago. The groups as defined in Arlequin (Schneider *et al.* 2000) for this dataset were based on political subdivisions within the Mongolian nation-state. Vigilant *et al.* (1991) provide two datasets with a worldwide distribution representing the HVRI and HVRII regions of the mitochondrial genome. The Kolman *et al.* HVRI samples ranged from positions 16020 to 16400, while the Ward *et al.* (1991) data represented positions 16021 to 16382.

The use of a worldwide dataset provided perspective on the multilayered structure of *H. sapiens* and helped place this species into the wider scheme of organismal evolution. The Vigilant HVRI and HVRII sequences ranged from 16024 to 16408 in the HVRI and from one to 408 in the HVRII. For the HVRI locus, 48 sequences with a worldwide range were used to represent all the haplotypes found in the dataset. The same methodology was used to reduce the HVRII dataset to 34 sequences. All human sequences were referenced against and rooted with Anderson *et al.* (1982), whose HVRI and HVRII sequences both originate from native British populations.

Among the teleost samples, Johnson (2000) provided d-loop samples from 12 dispersed populations of *Brachyrhaphis rhabdophora* (Costa Rican livebearing fish) occupying a series of streams bordering the Gulf of Nicoya in northwestern Costa Rica. While these fish are generally isolated by the extent of individual river basins, members of this species have been known to traverse the Gulf of Nicoya.

The Gondwanan distribution of *Galaxius maculatus* was sequenced by Waters and Burridge (1999). They provided one set of sequences from the Cytochrome b, and another set of sequences from the 16SrRNA region of the mitochondrial genome representing South America, Australasia, and Falkland Island populations. Before 1967, this species existed as 18 distinct and described species. However, McDowall (1967, in Waters and Burridge 1999) revised this group of organisms by lumping all populations into the species *Galaxias maculatus* and interpreting morphological differences as a consequence of ecological variation rather than genetic isolation.

The Brunner *et al.* (2001) dataset consisted of the d-loop taken from various Arctic Charr (*Salvelinus alpinus*) subspecific populations exhibiting a circumpolar distribution. In the analyses performed by Waters and Burridge (1999) and Brunner *et al.* (2001), phylogenetic and neighbor-joining trees produced groupings which closely mirrored geographical distributions. In both the original analysis and in this thesis, the dataset of Brunner *et al.* (2001) was rooted using a single *Salvenius svetovidovi* haplotype as an outgroup, while the Cytochrome b tree of Waters and Burridge (1999) used *Galaxias truttaceus* as an outgroup. Table 3 provided a description of each dataset used in this analysis.

While the control region in humans is highly variable (Salas *et al.* 2000), Lee *et al.* (1997) found that despite its rapid rate of evolution, the d-loop in fishes has less potential to carry phylogenetic information than it does in humans. Among human populations, the d-loop generally evolves at a much faster rate than the rest of the mitochondrial genome. Among other vertebrates, the rate of control region evolution is much closer to that of other genes in the mitochondria (Moritz *et al.* 1987). For a complete and detailed map of the mitochondrial genome, see Johnson *et al.* (1982).

**Table 3:** Description of datasets used in molecular analysis.

| Dataset | Organism and Range | Sample Size | Bases | Region of Mitochondria | Evolutionary Trend |
|---|---|---|---|---|---|
| Kolman *et al.* (1996) | *H. sapiens* (Mongolia) | 103 | 380 | HVR I | Multiethnic population-high gene flow |
| Ward *et al.* (1991) | *H. sapiens* (Nuu-Chah-Nulth) | 63 | 361 | HVR I | Single ethnic group-limited gene flow |
| Vigilant *et al.* (1991) | *H. sapiens* (Worldwide) | 188 | 384* | HVR I | Worldwide distribution-moderate gene flow |
| Vigilant *et al.* (1991) | *H. sapiens* (Worldwide) | 187 | 408* | HVR II | Worldwide distribution-moderate gene flow |
| Ovchinnikov *et al.*(2000) | Neanderthal (Caucasus) | 1 | 384* | HVR I | Ancient DNA |
| Krings *et al.* (1997), (2000) | Neanderthal (Germany and Croatia) | 2 | 408* | HVR I-HVR II | Ancient DNA |
| Johnson (2001) | *B. rhabdophora* (Northwestern Costa Rica) | 12 | 444 | D-loop | Local population-low gene flow |
| Waters and Burridge (1999) | *G. maculatus* (Gondwonan) | 6 | 520 | 16S | Gondwanan distribution-isolated local populations |

| Waters and Burridge (1999) | *G. maculatus* (Gondwonan) | 12 | 402 | Cytochrome b | Gondwanan distribution-isolated local populations |
|---|---|---|---|---|---|
| Brunner *et al.* (2001) | *S. alpinus* (Holarctic) | 62 | 552 | D-loop | Holarctic distribution-low gene flow between regional populations |

*- sequences of variable length

Through the examination of mitochondrial DNA, Ovchinnikov *et al.* (2000) estimate the divergence of neanderthal populations from the *H. sapiens* line at between 317,000 and 741,000 years ago. Krings *et al.* (1997) provide a more constant date of divergence at 500,000 years ago. In addition, Middle Eastern forms of neanderthal differ in terms of morphology from European forms. neanderthal samples into the HVRI and HVRII databases in order to test assumptions about the temporal dynamics of *H. sapiens* as a panmictic species and the place of neanderthal specimens within the genus *Homo*. Sampling the HVRI locus, Ovchinnikov *et al.* (2000) provide us with one sequence from a 29,000 year old specimen found in Mezmaiskaya Cave in the Caucasuses region of Western Asia, while Krings *et al.* (1997) provide two additional sequences from central Europe. In the case of the HVRII locus, Krings *et al.* (1997) and Krings *et al.* (2000) offer two sequences from central Europe and Croatia respectively. The Krings *et al.* (2000) HVRI and HVRII samples were taken from a 42,000 year old specimen found in Vindija Cave, Croatia, and a 40 to 100,000 year old specimen found in Feldhofer Cave, Germany. These neanderthal sequences were also been used to root each of the Vigilant *et al.* (1991) trees.

**Introduction to Subdivision within Molecular Datasets**

All eight datasets were partitioned into sample populations and groups on the basis of either geographic or systematic relationships. Among human datasets, the data was partitioned on the basis of ethnic groups and their geographic location. For example, in the Vigilant *et al.* (1991) dataset, all haplotypes representing the !Kung were placed in a certain sample and so were representative of that population at the frequency of which they occur in that population. While !Kung and Hadza populations represent different samples, they belonged to the same African group. These schemas were used in both the AMOVA and phylogenetic analyses.

In cases where there was limited ethnic variation such as in the Ward *et al.* (1991) sample, the dataset was randomly divided into samples that only allowed for a single group. Among the fish datasets, all samples in each dataset were divided up strictly on the basis of geography. For example, the Waters and Burridge (1999) dataset consisted of a single group formed by all South American and Falkland Island sample populations. Likewise, the Brunner *et al.* (2001) dataset was partitioned on the basis of regions in the North Atlantic. Table 4 shows the number of haplotypes representing each local population for the Vigilant *et al.* (1991) and Brunner *et al.* (2001) species-wide datasets, both of which involved multiple regional populations.

Among datasets where only a single group was allowed, the total amount of variance was restricted to the within group and within population levels. Even though the between group variation was not recorded in such cases, it was not found to affect comparisons made with datasets where all three levels of variance were recorded. In addition, data drawn from prior taxonomic investigations which suggested the integrity of relationships between samples were taken into account to produce populations and groups within Arlequin (Schneider *et al.* 2000). This was done in designing analyses of the Brunner *et al.* (2001) and Waters and Burridge (1999) datasets.

**Table 4:** Species-wide studies that involve more than two regional populations.

| Vigilant *et al.* (1991) | HVRI (n)* | HVR II (n)* | Vigilant *et al.* (1991) | HVRI (n)* | HVR II (n)* | Brunner *et al.* (2001) | D-loop (n)* |
|---|---|---|---|---|---|---|---|
| Polynesia | 1 | 0 | Indonesia | 1 | 1 | Atlantic | 18 |
| Australia | 2 | 0 | European | 5 | 2 | Acadian | 8 |
| Melanesia | 2 | 0 | Amerindian | 1 | 1 | Siberian | 9 |
| Pygmy | 7 | 8 | Hmong | 1 | 0 | Arctic | 18 |
| African-American | 1 | 3 | Chinese | 4 | 3 | Beringia | 9 |
| !Kung | 5 | 3 | Herero | 2 | 1 | | |
| Yoruba | 2 | 2 | Hadza | 3 | 2 | | |

*- number of haplotypes per population

**Reanalysis of Intraspecific Craniometric Traits**

The Howells (1973) worldwide dataset was used to test both the nature and degree of intraspecific variability for a single complex of related traits. Howells (1973) has previously shown how the nasion angle, gnathic index, and zygomatic breadth varies among modern human populations. These traits are not only evolutionarily plastic within *H. sapiens*, but are also significantly different among different species in the genus *Homo*. If these traits act to subdivide a species, then they should exhibit some degree of consistent directional change based on geography. The measurements for each trait were given in millimeters for zygomatic breadth and degrees of arc for nasion angle, while the standard formula for gnathic index is used. If taken together, nasion angle and gnathic index are especially important indicators of overall facial prognathism. The number of individuals sampled from each ethnic group is shown in Table 5.

**Table 5:** Populations used to determine intraspecific morphological subdivision and number of specimens for each  (Howells 1973)**.**

| Population | N | Population | N | Population | N | Population | N |
|------------|----|------------|----|------------|----|------------|----|
| Arikara | 42 | Norse | 55 | Egypt | 58 | Andaman | 35 |
| Eskimo | 53 | Euston | 35 | Bushman | 37 | Berg | 56 |
| Peru | 43 | Swanport | 35 | Dogon | 47 | Tasmanian | 30 |
| Buriat | 55 | Tolai | 56 | Zulu | 55 | Mokapu | 51 |
| Teita | 33 | | | | | | |

**Table 6:** Indices that transform continuous variables into discrete variables. Gnathic or Alveolar index (ratio of measures that indicate facial protrusion) = endobasion - prosthion length (27) x 100/ endobasion - nasion length (26).

| Classifications (Gnathic or Alveolar index) | |
|---------------------------------------------|---|
| Vertical face, orthogonal | Below 98 |
| Medium, mesognathous | 98-102.9 |
| Protruding face, prognathous | 103 or more |

By creating a grade scoring system, degrees of measures can be quantified using variables of form and extent, and as such become the most important criterion for delimiting patterns of diversity (Lahr 1996). The equation presented in Table 6 and taken from Rogers (1984) was used to define the gnathic index measurement in the Howells (1973) morphological dataset. This formula is also similar to how character states are delimited among Neotropical electric fish (Albert 1999).

**Methodology for Neotropical Electric Fish Analyses**

The species *G. carapo* was originally described by Linneaus in 1758, and is the most geographically widespread species in the genus *Gymnotus*. However, the holotype, or type specimen, is almost 250 years old and has not been georeferenced to a particular location. Compounding the problem of whether to call different looking specimens polymorphisms or new species is that there are only 11 species of *Gymnotus* currently recognized from southern Mexico to the Rio de la Plata in Argentina (Albert *et al.* 1999).

**Morphological Analysis of Neotropical Electric Fish**

Albert and Miller (1995) have provided published interspecific comparisons of anal-fin rays and pre-caudal vertebrae which were used to evaluate counts performed in the laboratory. The laboratory counts themselves were performed on 31 specimens representing five species of *Gymnotus*.

The eight major categories used to measure individual gymnotiform species were derived from Albert *et al.* (1999). These measurements were defined by the distance in millimeters between two specified morphological landmarks, and represent regions of the fish that have been found to be both plastic in evolution and conserved over time. All of the measurements referred to among *Gymnotus* specimens were shown in Table 7.

**Table 7:** Description of morphometric measurements used in this thesis.

| | |
|---|---|
| TL | Total body length, which is taken from the tip of the snout to the end of the tail. |
| HL | Head length of these individuals was taken from the tip of the snout to the back of the opercle. |
| PR | Pre-orbital length ranges from the tip of the snout to the anterior ridge of the eye orbit. |
| PO | Post-orbital measurement runs from the back of the opercle to the posterior ridge of the eye orbit. |
| BD | Body depth is measured from the anterior end of the anal fin straight up to the dorsal surface. |
| BW | Body width is taken by measuring the distance between the articulation of the left pectoral fin with the body and the articulation of the right pectoral fin with the body. |
| HD | Head depth, which is taken from the anus directly up to the top of the head. |
| MW | Mouth width is measured across the mouth from right to left. |

While a standard set of body measurements were collected on selected fish,

Albert *et al.* (1999) also reported a percentage which represents the comparison of head length (HL) against all measurements in the protocol except for head length itself (HL), body depth (BD), and body width (BW). HL, BD, and BW are compared to the total length (TL). This was done both for all known species with holotypes and established names, and groups of lots with a related morphology. All of these groups were then compared to one another.

**Statistical Analysis of Interspecific Diversity**

Specimens representing a specific point in space and time were defined as a lot. Individual lots were grouped with other such samples to define a species. For example, Albert (1999b) used two lots to represent *Gymnotus bahianus*, while also having used two lots from Brazil along with four lots from Venezuela to represent *Gymnotus cataniapo*. In this case, the diagnostic features in a species description were chosen broadly enough to mitigate the effects of most polymorphisms. When it was not absolutely clear which species a certain spatio-temporal sample belonged to, that lot was organized into a morphogroup. A morphogroup consists of several lots that possess morphological similarities. Using this methodology, it was predicted that statistical analysis would maintain the integrity of species-level groups. Thus, each measurement in the protocol was compared to the Head Length (HL). A bivariate graph was plotted to compare the two measurements. Each specimen represented a datum point, and the data points of each morphogroup were color-coded. Ideally, the members of each morphogroup should cluster along a regression line.

In bivariate comparisons of interspecific growth rates, head length was the independent variable which varied only with growth. Head length was plotted along the abscissa, and was contrasted with a series of dependent variables plotted on the y-axis. The dependent variable was plotted along the y-axis because it helped provide insight into controlled changes of the evolutionarily stable independent variable. Graphical comparisons were used to examine three variables in this manner, two of which were known to be evolutionarily plastic and one which was considered conserved. These graphs also tested whether or not the rates of growth among pre-orbital, mouth width, and post-orbital measurements provided evidence for taxonomic distance between groups of specimens. Thus, when the trend of development among different groups followed a diverging trajectory, it provided evidence that a certain characters can aid in distinguishing between taxa.

# RESULTS

All datasets analyzed here were chosen on the basis of their potential to express both typical and interesting trends in human and fish evolution. The Johnson (2001), Kolman *et al.* (1996), and Ward *et al.* (1991) molecular samples were chosen to simulate the nature of local intraspecific populations. By contrast, the Vigilant *et al.* (1991) datasets were chosen to provide a species-wide look at human molecular diversity. To test for any potential intraspecific discontinuity in the human dataset, the two fish species originally sampled by Brunner *et al.* (2001) and Waters and Burridge (1999) were chosen. These teleost examples have been found to exhibit significant amounts of discontinuity in previous analyses.

The neanderthal samples were chosen to test for the existence of an interspecific relationship between themselves and modern humans when tested against molecular data from fish. This thesis relied exclusively upon molecular evidence for reaching conclusions regarding the possibility of interbreeding and the existence of taxonomic distance between modern human and neanderthal populations. The gymnotiform specimens provided a teleost interspecific morphological analogue for discussing ways to distinguish between taxonomic groups and the process of heterochrony within the genus *Homo*.

The findings presented here cover analyses from several previously published studies. For purposes of facilitating comparisons between these datasets, each molecular dataset is referred to in the text with an abbreviated suffix. All datasets representing *H. sapiens* were assigned the symbol (h), while all of the fish datasets were referred to by the symbol (nh).

**Results of Analyzing Variance Among Molecular Datasets**

No clear genetic discontinuity was found to exist among contemporary *H. sapiens* populations. Even in cases where the human species has colonized new continents to expand its geographic range, there has not been the same degree of geographical subdivision as seen among some fish species. The results of AMOVA analyses performed on all four modern human datasets provided quantitative evidence supporting these contentions. Table 8 provided a hierarchical breakdown of variance and differentiation between various populations of *H. sapiens* and teleosts.

The results of these AMOVA analyses provide quantitative evidence for several trends which transcend the differences between hominids and teleosts. It was found that for both local populations represented by Ward *et al.* (1991), Kolman *et al.* (1996) and Johnson (2001) and panmictic species represented by the Vigilant *et al.* (1991) datasets, there existed more diversity within local populations than between regional groups. This trend was most apparent within the Kolman *et al.* (h)(1996) and Ward *et al.* (h)(1991) HVRI datasets. Meanwhile, species with known subspecies or discontinuous geographic distributions exhibited large percentages of between group variance. For example, the Brunner *et al.* (nh)(2001) and both Waters and Burridge (nh)(1999) datasets exhibited more between group differentiation than for any other level of variance.

The neanderthal samples exhibited a unusually high degree of variance for what was presumed to be a single, panmictic species and relatively small sample size. This may be due to a thymine doublet which exists in the German sequence, but not the other two sequences in the dataset. A difference of four mutational steps in a dataset of three taxa can result in such large amounts of intergroup diversity in a single species.

When the thymine doublet was removed and the AMOVA analysis was run again, a lower but still abnormally high amount of within population variance. Such results were found to contrast with the Johnson (nh)(2001) sample, in which 12 samples were separated by only three mutational steps while also exhibiting a much lower within group variance.

The Johnson (nh)(2000) samples can also be directly compared to the Ward (h) (1991) samples for purposes of demonstrating levels of diversity within a small population. Among Costa Rican Livebearing Fish, Johnson (nh)(2001) found that while there was extensive gene flow within the stream networks of northwestern Costa Rica, there was also a hierarchical network of rivers and streams which structures the entire species range. This may have allowed for extensive genetic drift between local populations in this species given the proper amount of time and a lack of rapid dispersal events.

The AMOVA (Schneider *et al.* 2000) and PAUP (Swofford 1993) analyses of the Vigilant *et al.* (h)(1991) datatsets used in this thesis were designed in part to detect signals of geographic differentiation in *H. sapiens*. Nei (1982) used blood proteins from four African, five European, and ten Asian populations to test whether or not subdivision exists within *H. sapiens*. According to these data, samples in the African group were found to have diverged from the rest of *H. sapiens* twice as long ago on average as the divergence point between Caucasian and Mongoloid samples. However, the between population and between group levels of variance for the Vigilant *et al.* (1991)(h) datasets were not large enough to produce sufficient evidence for the existence of distinct subdivisions.

**Table 8:** Comparison of AMOVA results between eight datasets, recorded as total percent variance of dataset.

| Dataset | Between Groups | Within Groups | Within Populations |
|---|---|---|---|
| Johnson (nh)(2001) Costa Rican Livebearing Fish | ------- | 16.67 | 83.33 |
| Waters and Burridge (nh)(1999) *G. maculatus* 16S | 50.8 | 4.2 | 44.99 |
| Waters and Burridge (nh)(1999) *G. maculatus* Cytochrome b | 72.44 | 0.46 | 27.11 |
| Brunner *et al.* (nh)(2001) Arctic Charr | 47.11 | 31.2 | 21.69 |
| Ward *et al.* (h)(1991) Nuu-Chah-Nulth | ------- | 15.99 | 84.01 |
| Kolman *et al.* (h)(1996) Mongolian | ------- | 1.13 | 98.87 |
| Vigilant *et al.* (h)(1991) HVRI worldwide human | 0.76 | 42.24 | 57.00 |
| Vigilant *et al.* (h)(1991) HVRII worldwide human | 7.04 | 26.34 | 66.62 |
| Vigilant *et al.* (h)(1991) HVRI human/neanderthal | 21.01 | 33.82 | 45.18 |
| Vigilant *et al.* (h)(1991) HVRII human/neanderthal | 30.6 | 17.48 | 51.92 |
| HVRI neanderthal only | ------- | 33.71 | 66.29 |
| HVRI neanderthal w/o thymine doublet | ------- | 16.52 | 83.48 |

This was especially true when compared to the intraspecific fish molecular datasets of Waters and Burridge (nh)(1999) and Brunner *et al.* (nh)(2001) shown in Table 8. In addition, the results of the AMOVA analyses on the Vigilant *et al.* (1991) datasets were also found to correlate with other AMOVA analyses of a worldwide distribution of humans.

**Results of the Molecular Phylogenies**

Several phylogenetic trees were produced using PAUP (Swofford 1994) based on the eight datasets analyzed in Arlequin (Schneider *et al.* 2000). Of these, the Waters and Burridge (nh)(1999), Vigilant *et al.* (h)(1991), and Brunner *et al.* (nh) (2001) phylogenies were found to be most relevant for looking at variations within species-level tree topologies. Intraspecific discontinuity was gauged by observing both monophyly among regionally exclusive populations which was also supported by high bootstrap values and by looking at the number of steps separating resolved clades from the rest of the dataset. In the figures showing tree topologies, both bootstrap values and the number of steps along selected branch lengths are listed.

In terms of the demic-level populations represented by Ward (h)(1991), Kolman *et al.* (h)(1996), and Johnson (nh)(2001), tree topologies produced by PAUP (Swofford 1994) showed little important internal differentiation. The lack of structuring reached a nadir in Johnson (nh)(2001). While two distinct clades were found to exist, each of them yielded internal polytomies. The structuring that did exist was found to be correlated with the existence of different haplotypes present in a single population. In fact, when dealing exclusively at the level of small populations, not much differentiation between samples should be expected. For example, a bootstrap analysis of the 12 Costa Rican livebearing fish populations sampled by Johnson (nh)(2001) demonstrated a reasonable amount of structuring considering the lack of variance among the samples analyzed. These populations formed only three haplotypes distributed geographically in such a way so that two haplotypes existed exclusively among four northern groups, while a third haplotype was shared by eight southern groups and one northern group.

This situation was reflected in the bootstrap analysis, which produced a single clade with a score of 73 % as shown in Figure 1. This may have been the result of either the fixation of a single optimal haplotype by means of some advantageous maternally-inherited factor, or the occurrence of a recent and severe genetic bottleneck (Johnson (nh) 2001). In keeping with the original conclusion reached by Johnson (nh)(2001), there was found to be a general restriction on fecund females in northwestern Costa Rica sometime in the recent past.

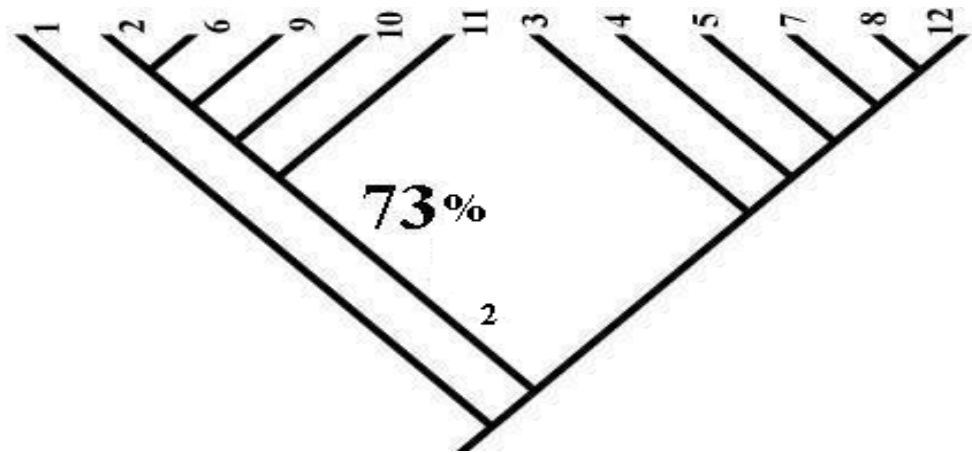

**Figure 1:** Bootstrap analysis based on a heuristic search of the Johnson (nh)(2001) dataset (bootstrap scores under 60% not shown).

The topology of the Kolman *et al.* (h)(1996) bootstrap analysis shown in Figure 2 exhibited no clearly defined monophyletic groupings. Similarly, the Ward *et al.* (h) (1991) bootstrap analysis shown in Figure 3 only exhibited a single clearly defined subgroup of five sequences supported by a 62 % bootstrap score.

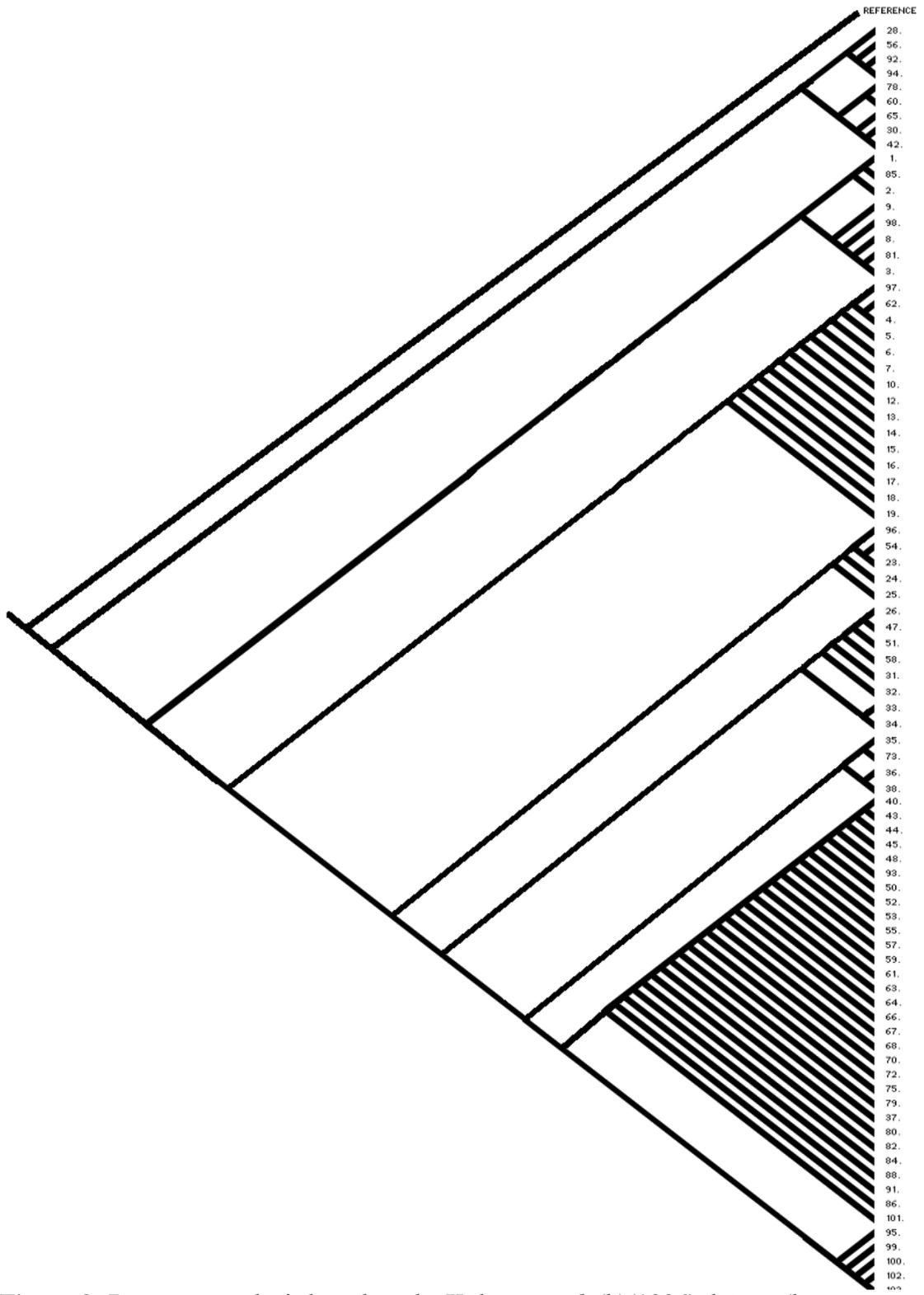

**Figure 2:** Bootstrap analysis based on the Kolman *et al.* (h)(1996) dataset (bootstrap scores under 60% not shown).

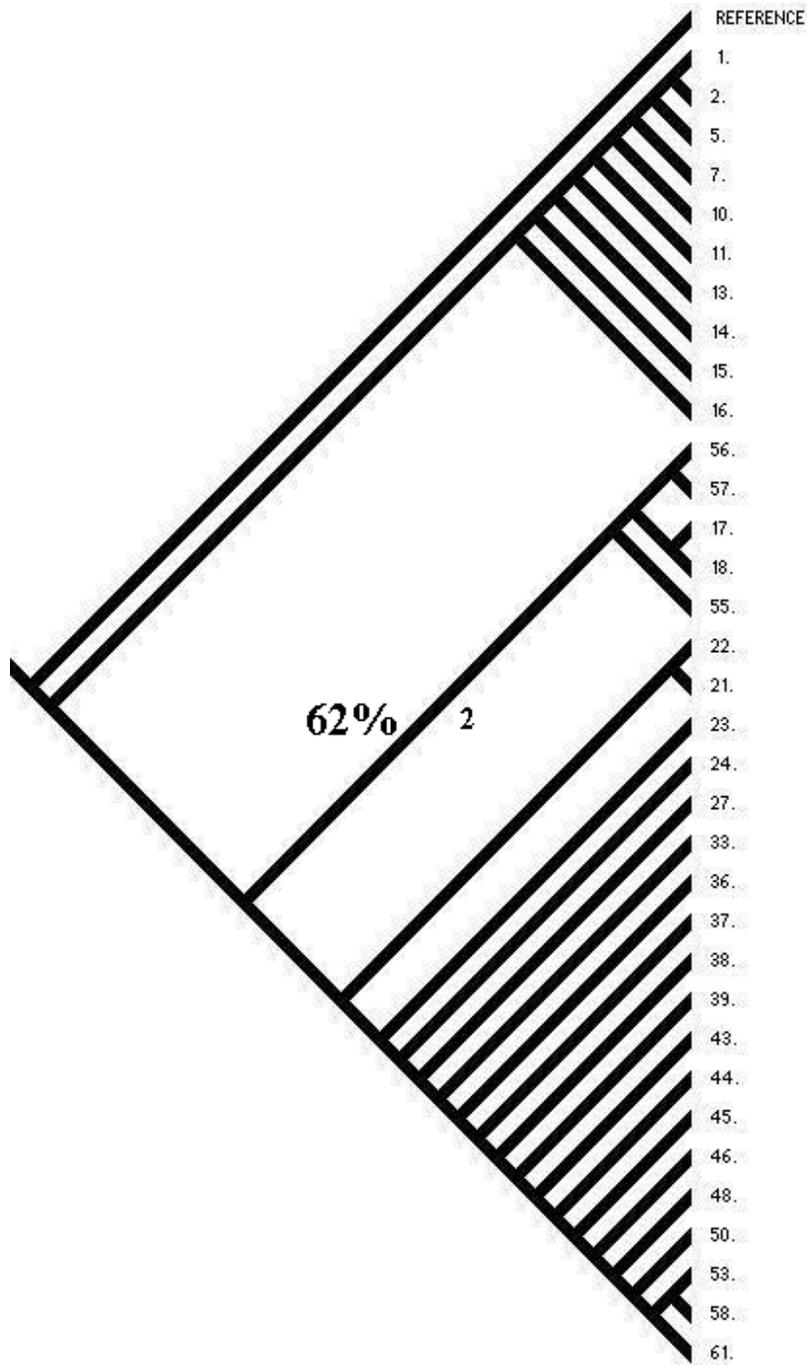

**Figure 3:** Bootstrap analysis based on the Ward *et al.* (h)(1991) dataset (bootstrap scores under 60% not shown).

This lack of significant bootstrap scores and long branch lengths demonstrate the weak degree of structuring in these topologies. All haplotypes featured in the Kolman *et al.* (h)(1996) and Ward *et al.* (h)(1991) analyses were numbered according to the scheme used by HVRbase. This database was found on the World Wide Web at www.HVRbase.de, and provided descriptions of representatives for each haplotype which corresponded to their assigned number. Any differences between the trees produced by Ward *et al.* (h)(1991) and the topologies presented here may be due to either differences in the version of PAUP used (Swofford 1993) or the overall computing power available for either analysis.

The general lack of within group differentiation found among Mongolians and Nuu-Chah-Nulth populations was also reproduced when both groups were compared to each other. Due to the consistent lack of phylogenetic structuring in these local populations, little to no phylogenetic discontinuity was interpreted more generally as what characterized a lack of internal subdivision within larger species-level populations. While it was recognized that smaller local populations do not exhibit panmixis in the same way species-level populations do, phylogenetic structuring was used as more of a heuristic tool which acted to complement the findings of the AMOVA analyses.

By contrast, the Brunner *et al.* (nh)(2001) dataset shown in Figure 4 exhibited extensive internal discontinuity. The bootstrap analysis shown in Figure 4 yielded single clade consisting of 18 Atlantic specimens was supported by a 62 % bootstrap score. Likewise, 18 Arctic specimens formed a clade supported by a 96 % bootstrap score. A third distinct clade was formed by eight Acadian specimens and was supported by a 95 % bootstrap score. These three populations were also distinct from 18 sequences representing specimens from Siberia and Beringia. This stands in sharp contrast to what was found among the Vigilant *et al.* (h) (1991) datasets. Both datasets were rooted with the Anderson *et al.* (1982) modern human reference sequence and non-human primate sequences. For the HVRI tree topology shown in Figure 5, two distinct clades were produced.

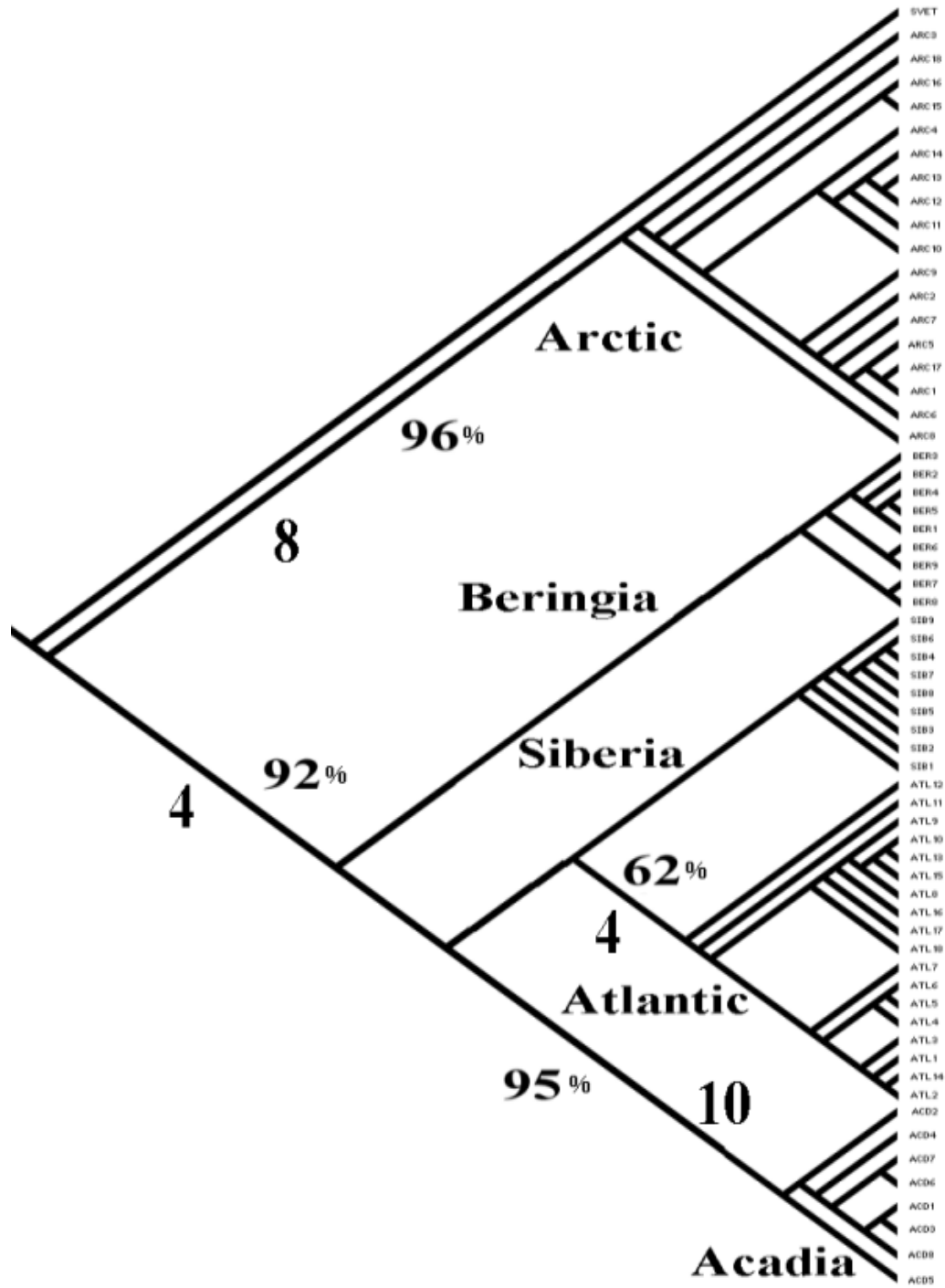

**Figure 4:** Results for the Brunner *et al.* (nh)(2001) bootstrap analysis (bootstrap scores under 60% not shown).

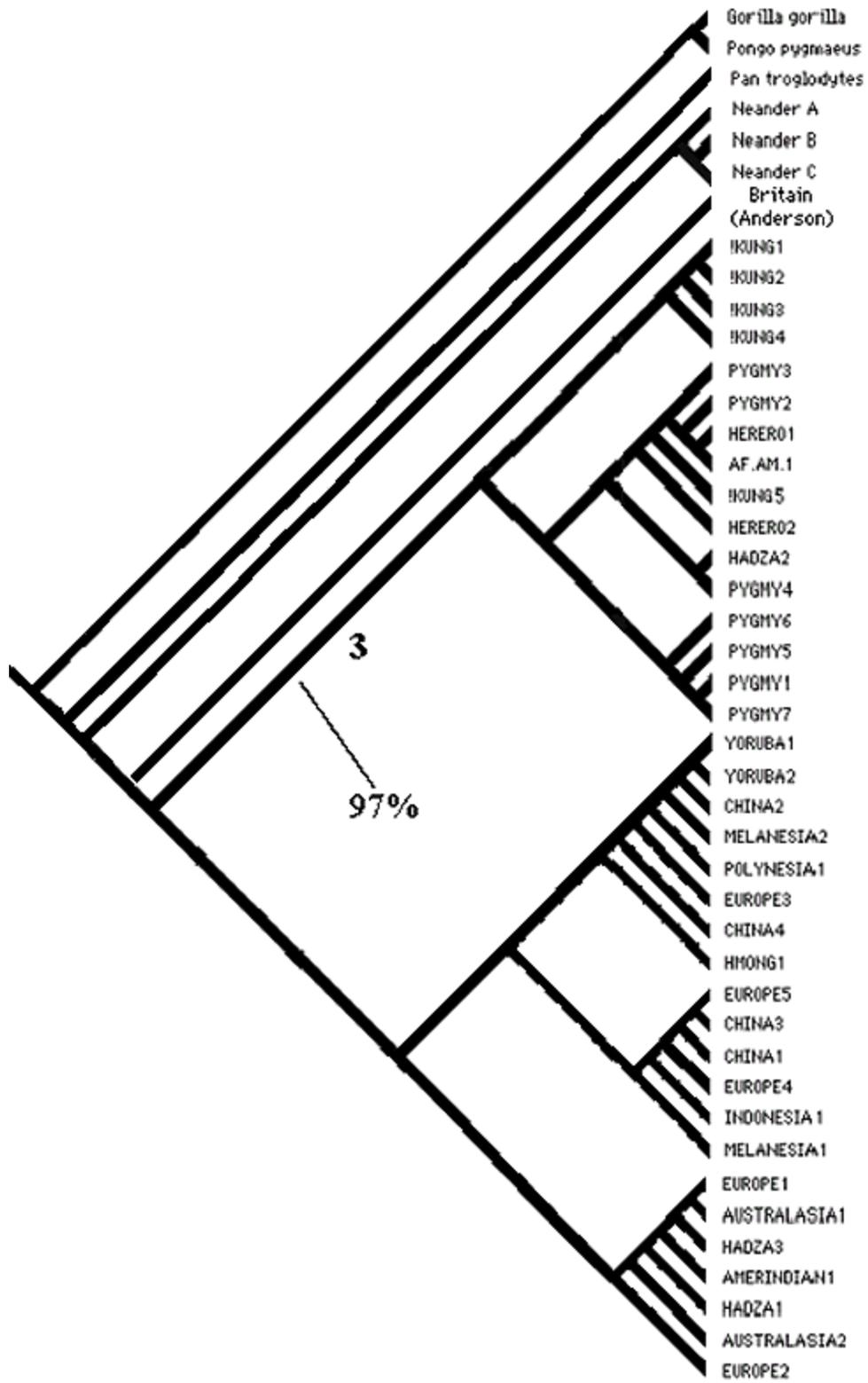

**Figure 5:** Results for the Vigilant *et al.* (h)(1991) HVRI bootstrap analysis involving a worldwide distribution of modern humans (bootstrap scores under 60% not shown).

One clade in the HVRI topology was entirely African in origin, supported by an extremely significant bootstrap score of 97%. While this seemed to indicate intraspecific discontinuity within *H. sapiens*, the other clade in the topology yielded both African and non-African haplotypes. Therefore, while subdivision was found to exist within the HVRI dataset, it was not clearly based on distinct geographical units as was the case among the Brunner *et al.* (2001) Arctic Charr dataset.

The HVRII dataset shown in Figure 6 failed to exhibit any clades with significant bootstrap support. In addition, while regional populations were generally found near other populations from the same region, there was little phylogenetic resolution throughout the tree topology. As a consequence, little discontinuity was demonstrated in this dataset.

The subdivisions found among the Brunner *et al.* (2001) Arctic Charr samples shown in Figure 4, however, were not of a strictly discontinuous nature as was found in phylogenies based on the species *G. maculatus*. Bootstrap analyses performed on the Waters and Burridge (nh) (1999) datasets shown in Figures 7 and 8 provide additional quantitative support for the discontinuous nature of this species. The 16S analysis shown in Figure 7 yielded only one resolvable clade. Yet this clade consisted of all South American and Falkland Islands specimens, and was supported by a 100% bootstrap score.

Meanwhile, the Cytochrome b analysis shown in Figure 8 yielded two distinct clades each with bootstrap support of 99 % each. One clade consisted of South American specimens, and the other consisted of Australasian specimens.

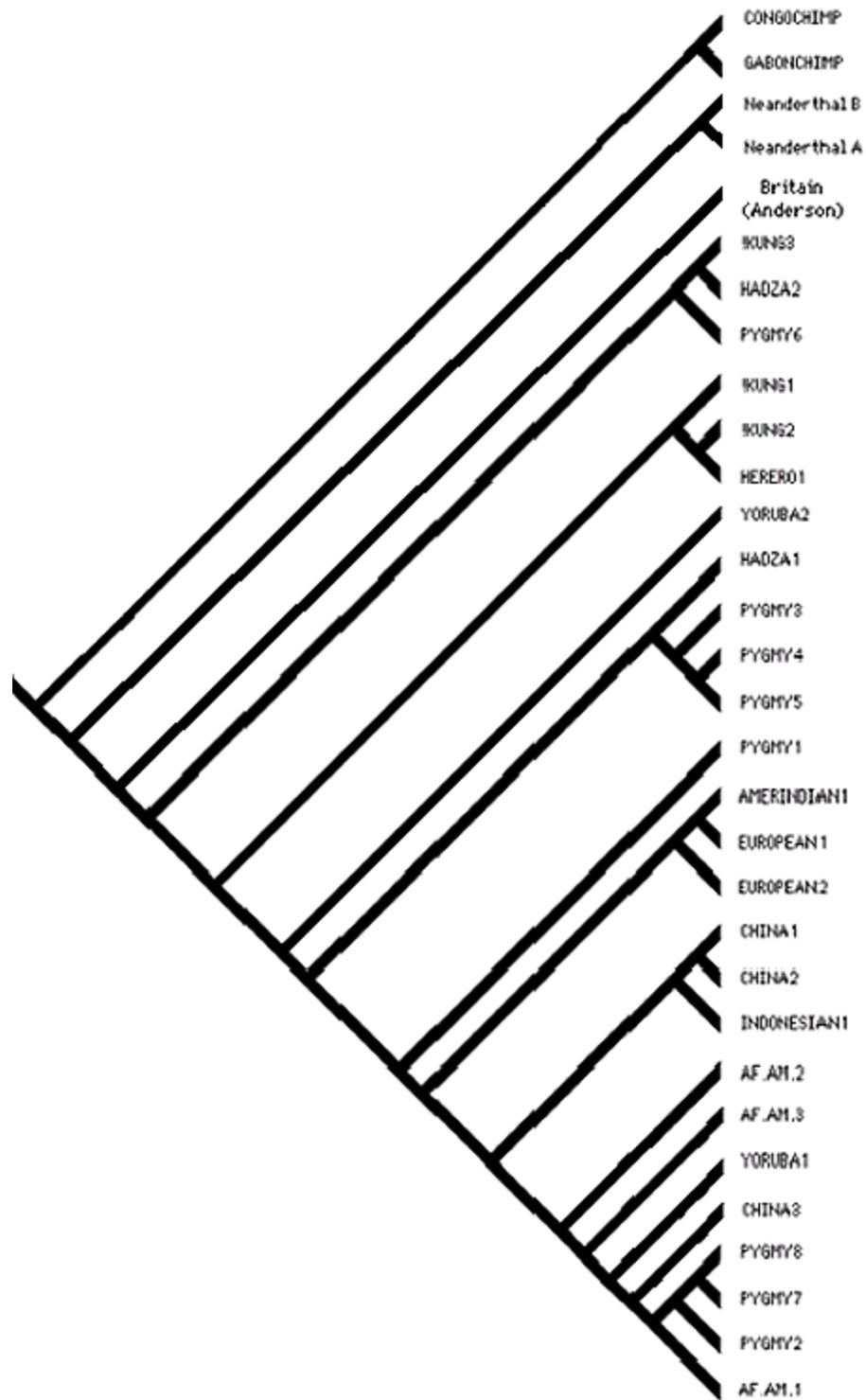

**Figure 6:** Results for the Vigilant *et al.* (h)(1991) HVRII bootstrap analysis involving a worldwide distribution of modern humans (bootstrap scores under 60% not shown).

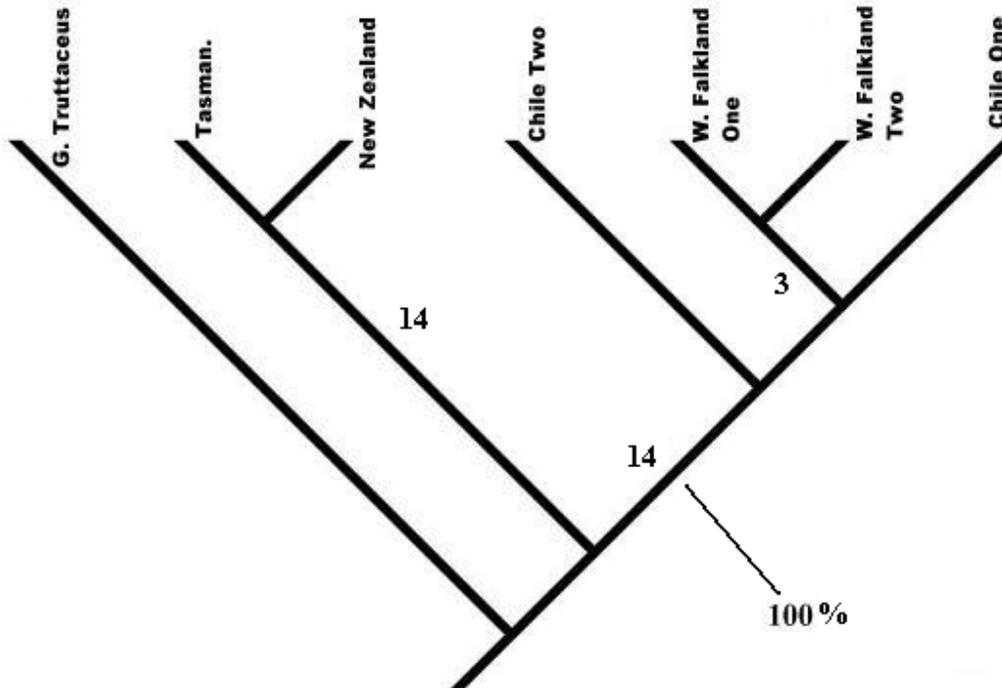

**Figure 7:** Topology resulting from a bootstrap analysis of the *G. maculatus* 16S dataset (bootstrap scores under 60% not shown).

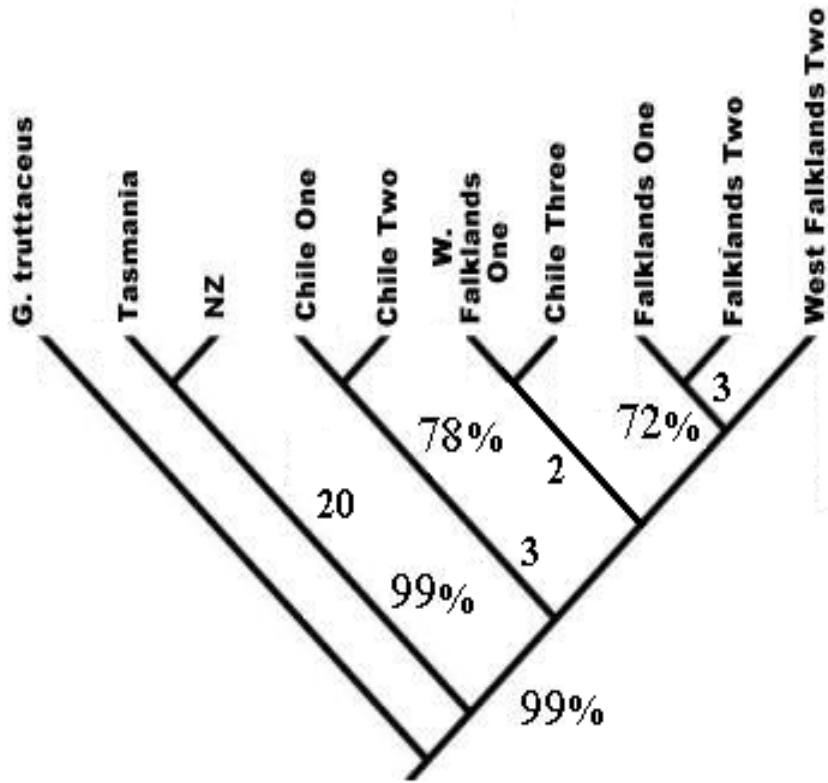

**Figure 8:** Topology resulting from a bootstrap analysis of the *G. maculatus* Cytochrome b dataset (bootstrap scores under 60% not shown).

Within the South American group, two clades with bootstrap scores of 78 and 72 percent were found. These clades were not subdivided by geography, and both included specimens from Chile and the Falkland Islands. This evidence points to clear subdivision within *G. maculatus* based on geography.

Finally, a formal test was performed to determine whether or not the highly supported African clade in the Vigilant *et al.* (h)(1991) bootstrap topology represents intraspecific discontinuity much like highly supported clades do among regionally distinct *G. maculatus* specimens. Using the "trace characters" option in MacClade 3.1 (Maddison and Maddison 1993), evolutionary changes for each nucleotide base were observed. When a nucleotide showed a change in state exclusively within each well-defined clade described in Table 9, it was recorded and compared to the total length of the sequence.

**Table 9:** Number of nucleotide bases which define discontinuity supported by selected bootstrap analyses.

| Dataset | Length of Sequence (bases) | Clade Observed and its Bootstrap Support | Unique Polymorphisms | Percent of Total Sequence |
|---------|---------|---------|---------|---------|
| Vigilant *et al.* (h)(1991) HVRI | 384 | African (97) | 3 | 0.007 |
| Waters and Burridge (nh)(1999) 16S | 519 | Magellan (100) | 14 | 0.03 |
| Waters and Burridge (nh)(1999) Cytochrome b | 402 | Australasian (99) | 20 | 0.05 |

The result of this test showed that the percentage of polymorphisms unique monophyletic groups supported by high bootstrap values as compared to the entire sequence was nearly an order of magnitude higher among the *G. maculatus* datasets. It was also of note that the branch lengths separating the clades within the Magellan group shown in Figure 8 were comparable to the extent of discontinuity within *H. sapiens*. By contrast, the branch lengths that separate subgroups within *S. alpinus* were several steps longer than any discontinuity found within the Vigilant *et al.* (h)(1991) dataset.

**Intraspecific Morphological Variation Within *Homo sapiens***

The amount of between group variance in the species *S. alpinus* was found to be greater than differentiation at the within group and within population levels. Coupling this observation with data on geographic distribution and habitat, many examples of previously established intraspecific discontinuities within the species *S. alpinus* were found to be monophyletic. Contrasting this result with the geographically and regionally unstructured Vigilant *et al.* (h)(1991) worldwide datasets, it was apparent that there is little systematic evidence for intraspecific discontinuity within the range of modern human diversity using molecular data.

Discontinuous intraspecific diversity is a natural feature of animal biodiversity produced by the same forces that ultimately distinguish between species. The parameters of discontinuity within a species has been established through taxonomic distance analyses that yield geographically-based groups separated by a total mean sequence divergence of one half percent or greater (Avise *et al.* 1987). In addition, Lake Cisco (*Coregonus artedi*) and many other species of northern temperate freshwater fish exhibit subpopulations known as glacial races. Such intraspecific groups have been found to be products of a geographic restriction of these populations to refugia during the Pleistocene (Turgeon and Bernatchez 2001). Yet not only should caution be taken in correlating these findings with what is found within *H. sapiens,* but morphological data from modern humans must also be investigated to make any proposal of discontinuity within modern humans.

The measurements for three morphological characters involved with facial prognathism among seventeen populations were taken directly from Howells (1973) in an attempt to compare the lack of molecular discontinuity with what is found among morphological data. The mean and standard deviations for these populations divided on the basis of five geographic regions were then calculated for purposes of this thesis. The measurements taken by Howells (1973) were then used to understand trends produced by the regional averages. It has been shown that throughout the species *H. sapiens,* there was a general trend for the Gnathic Index to increase as the Nasion Angle did the same. However, no clear relationship has been found between any sort of trend in size and geographic distribution. This conclusion was based on comparing standard deviations for each of the five continents in Table 10.

It was found that specimens from Swanport in England exhibited a high Gnathic Index, along with specimens from Tasmania and Sub-Saharan Africa. Likewise, European specimens like the Norse and Berg samples have been found to be comparable with Eskimo and Egyptian samples for this trait. When the means and standard deviations were calculated and compared, none of the aggregate means for each continent were found to be statistically significant from one another.

The standard deviations was used to represent regional variation among each trait. In terms of the zygomatic arch, however, there was found to be a general increase in size among Asian, Amerindian, and Oceania specimens. These populations were also statistically significant on average from the African and European specimens. When the characteristics in this analysis were taken collectively, *H. sapiens* could only be subdivided into distinct groups by using an extremely liberal criteria. Overall, these traits do not seem to be exclusive to any one geographic area.

In terms of looking at modern human morphological characters over space and time, a more comprehensive study of 23 craniofacial measurements from 1,802 *H. sapiens* crania was performed by Hanihara (1996). When regional populations were compared, several discontinuities between the clinal nature of traits and geographic distribution appeared. For example, Australians were shown to have a closer affinity to Africans than to Polynesians. Furthermore, recent Europeans were found to be similar to East Asians while among East Asians themselves there were differences between northern and southern Asian populations. All this evidence points to the idea that the evolutionary divergence of craniofacial shape may be a recent phenomena. Furthermore, the existence of such patterns in variation make it difficult to reconstruct evolutionary relationships among regional populations.

Local populations of *H. sapiens* subdivided on the basis of these craniometric traits tend not to correlate with geographic partitions. As was also be seen among Neotropical electric fishes, traits must exhibit more than wide variation. Plastic traits must follow consistent trends towards change in size and form over time.

**Table 10:** Measures of facial prognathism in males from modern living populations.

| Population | Gnathic Index | Nasion Angle | Zygomatic Breadth |
|---|---|---|---|
| Eskimo* | 96.8 | 67.5 | 100 |
| Americas Mean** | 96.97 | 67.1 | 99.3 |
| Americas Standard Deviation** | 1.16 | 0.96 | 2.1 |
| Asia Mean** | 98.2 | 68.25 | 98.5 |
| Asia Standard Deviation** | 1.41 | 2.9 | 7.8 |
| Berg* | 95.1 | 65.6 | 93 |
| Norse* | 95.2 | 66.1 | 94 |
| Swanport* | 103.5 | 74.8 | -- |
| Europe Mean** | 99.3 | 70.08 | 93.5 |
| Europe Standard Deviation** | 4.79 | 4.9 | 0.7 |
| Tasmanian* | 101.6 | 75.3 | 95 |
| Oceania Mean** | 101.57 | 73.5 | 97.3 |
| Oceania Standard Deviation** | 4.05 | 4.01 | 2.1 |
| Egypt* | 95.1 | 65.9 | 94 |
| Bushman* | 98.8 | 71.2 | 92 |
| Dogon* | 101.4 | 72.1 | 96 |
| Zulu* | 100.4 | 71.1 | 96 |
| Teita* | 100.2 | 71.3 | 99 |
| Africa Mean** | 99.18 | 70.32 | 95.4 |
| Africa Standard Deviation** | 2.46 | 2.5 | 2.6 |

**\*- from Howells (1973), \*\*- from this thesis**

In addition, intraspecific groups which differ from one another on the basis of discrete traits must also be confined to specific and contiguous geographic units. Therefore, while craniofacial traits were shown to be plastic within *H. sapiens* they appear to be too variable in terms of defining significant intraspecific discontinuity.

**Tests of Limited Inbreeding Between Neanderthal and *Homo sapiens***

**Cladistic Exchange Experiment**

Using the Vigilant *et al.* (h)(1991) HVRI and HVRII datasets and neanderthal sequences from Krings *et al.* (1997) and Ovchinnikov *et al.* (2000), the cladistic exchange experiment was performed to detect limited inbreeding between neanderthal and modern humans. Given the evolutionary constraints described by Huelsenbeck (1991), this test involved using a neanderthal clade highly supported by bootstrap values to perform one of three operations. The first operation involved placing a single neanderthal clade in between a number of different clades within the range of modern human diversity. The second operation placed a number of clearly definable modern human clades within the neanderthal clade one at a time. The third operation consisted of leaving the neanderthal clade in place while modern human clades were exchanged with one another. The movement of clades and individual branches across the range of diversity for a dataset were similar to the operations used by search algorithms to define a most parsimonious tree. These manipulations of an already established most parsimonious tree were performed using MacClade 3.1 (Maddison and Maddison 1993). Their purpose was to simulate how many evolutionary steps would be added to a most parsimonious tree if neanderthal haplotypes were considered to be part of the total range of human molecular diversity.

It was found that only a few extra steps separates the neanderthal exchanges into modern humans from instances where segments of modern human diversity were rearranged relative to the total number of steps in the tree. The most interesting result of this exercise seems to be that when neanderthal specimens were placed in between clades within the range of modern *H. sapiens*, the number of steps greater than the most parsimonious tree only increased by five to eight steps. This result was contingent upon the exact placement of the neanderthal clade. This was opposed to what happened when individual monophyletic clades from the range of human diversity were placed within the neanderthal clade for both of the Vigilant *et al.* (h)(1991) HVRI and HVRII phylogenies. In these cases, the number of extra steps ranged from 10 to 28. This may partially be due to each neanderthal clade being supported by a bootstrap score of 100, as the tree statistics hardly changed from manipulation to manipulation. Table 11 shows the full results of this exercise.

The exact quantitative nature of this phenomenon was more effectively modeled in the experimental exercise by placing the neanderthal sequences into the range of modern humans. While the following results show that many more steps were added with the inclusion of modern humans into neanderthal, they do not on their own explain what this means taxonomically. Generally, each taxonomic unit exhibits a range of diversity measured in mutational steps which was discontinuous with other such populations. While this does not guarantee distinction between taxonomic groups, it should be assessed and compared with populations known to be closely related. For example, based on the following results it can be hypothesized that the range of human diversity varies up to three steps greater than the most parsimonious tree for each hypervariable region. This was supported by moving a single modern human clade to a point in between other clades in the Vigilant *et al.* (h)(1991) modern human phylogeny.

**Table 11:** Clades moved in and out of neanderthal and modern human populations (statistics for the original most parsimonious tree in parentheses).

| Operation Performed (HVRI) | Steps from MPT* (470) | Operation Performed (HVRII) | Steps from MPT* (346) |
|---|---|---|---|
| Nean between A and B- | 7 | Nean between A and B- | 5 |
| Nean between B and C- | 7 | Nean between B and C- | 7 |
| Nean between D and E- | 8 | Nean between C and D- | 7 |
| Nean between F and G- | 6 | Nean between D and E- | 5 |
| A into neanderthal | 28 | A into neanderthal | 11 |
| B into neanderthal | 21 | B into neanderthal | 7 |
| C into neanderthal | 22 | C into neanderthal | 10 |
| D into neanderthal | 21 | D into neanderthal | 10 |
| E into neanderthal | 23 | E into neanderthal | 15 |
| F into neanderthal | 19 | A between D and E- | 1 |
| G into neanderthal- | 22 | B between D and E- | 1 |
| A between B and C- | 2 | E between A and B- | 1 |
| G between B and C- | 3 | | |
| D between E and F- | 2 | | |

*- most parsimonious tree.

This experimental exercise also required the identification of several modern human groups taken from the Vigilant *et al.* (1991) global dataset. A total of seven clades in the HVRI and five clades in the HVRII were defined from the Vigilant *et al.* (1991) modern human datasets. The groups defined in Table 12 were interpositioned between each other along with the neanderthal specimens for each hypervariable region as shown in Figure 9.

**Table 12:** Definition of groups in the cladistic exchange experiment.

| Group Definitions (HVRI) | Group Definitions (HVRII) |
|---|---|
| A- Pygmy, European, and Amerindian | A- !Kung and China |
| B- !Kung and Australasia | B- Bantu and Herero |
| C- Hadza and Europe | C- Yoruban, Hadza, and European |
| D- Pygmy and East Asia | D- !Kung and Hadza |
| E- !Kung and Europe | E- Bantu and Hadza |
| F- Pygmy | |
| G- !Kung and East Asia | |

Making all possible exchanges among and between the seven clades from the HVRI and five groups from the HVRII analysis resulted in a maximum perturbation of three extra steps. By contrast, moving neanderthal into the range of human diversity results in an average increase of 6.5 steps. This difference was most likely due to specific-level differences between the neanderthal samples and modern human samples. Overall, this result supports the contention that the amount of interbreeding between modern humans and neanderthal was small. This result implies that the possibility exists for limited inbreeding between neanderthal and modern humans.

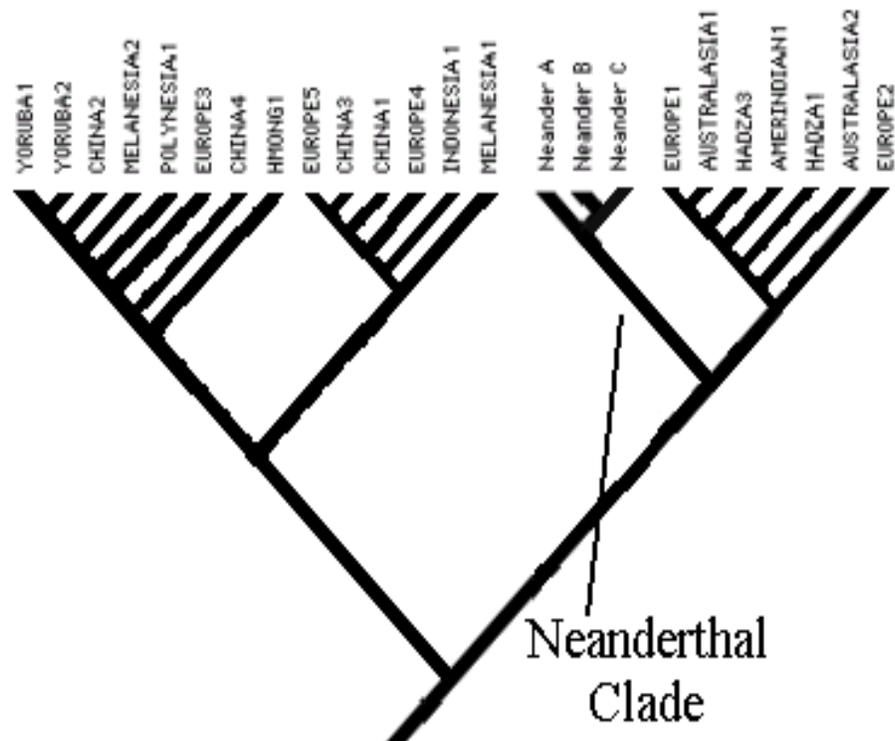

**Figure 9:** An example of how the neanderthal clade was interpositioned between two modern human clades using the HVRI dataset as an example.

These data showed that if *H. sapiens* and neanderthal were considered a single species without subspecific definitions, it would exhibit a discontinuous haplotype distribution separated by at least 10 mutational steps. A three to six percent divergence is comparable with what was observed within extant species of the genus *Lepomis* (Bermingham and Avise 1986). This provides some support for a limited area of backcrossing among ancient *H. sapiens* such as exists among local populations of *Lepomis macrochirus* in Georgia and South Carolina (Avise *et al.* 1987).

**Taxonomic Variance Test Involving Molecular Data**

An experimental exercise called the taxonomic variance test was performed that tested whether or not neanderthals and modern humans constitute independent taxonomic groups. The two main hypotheses concerning the taxonomic status of both populations was tested in Arlequin (Schneider *et al.* 2000) by using the Vigilant *et al.* (h)(1991) worldwide HVRI and HVRII data plus added neanderthal sequences.

The single species examples treated neanderthal as a regular regional population within the context of eight geographically defined groups of modern humans. This dataset then tested the hypothesis that modern humans and neanderthals were both polymorphic variants of the species *H. sapiens*. By contrast, the data file used to test the hypothesis that neanderthal was a subspecies of modern *H. sapiens* placed all geographically defined modern human populations in one group and all neanderthal sequences in a second group. The results of this exercise were shown in Table 13.

It was found that the percentage of variation both within and between groups was much higher when the neanderthal sequences were treated as a subspecies. It was of note that including neanderthal within the species group *H. sapiens* resulted in percentages of variation that were distorted from their values before the neanderthal sequences were introduced. This distortion also effectively argues against the inclusion of neanderthal into *H. sapiens*. On the other hand, treating them as separate species created a notable difference in the nature of the sample's population architecture. By contrast, when the neanderthal sequences were separated and treated as a group separate from modern humans, the between group variance was great while the within population variance was small. This was exactly the opposite of what was consistently shown for modern humans and the Johnson (nh)(2001) fish sample. Thus, it was found that treating neanderthal as a group separate from modern humans group resulted in a more informative expression of variance components as shown in Table 13.

Specimens of neanderthal were included in a cladistic analysis with all the haplotypes from the Vigilant *et al.* (h)(1991) HVRI dataset and rooted with sequences from three great ape species (Horai *et al.* 1992) and the Anderson *et al.* (1982) modern human reference sequence. Figure 10 graphically demonstrated that all three neanderthal samples from that analysisformed an clade independent of all modern human sequences.

**Table 13:** Hypotheses regarding taxonomic status between neanderthal and modern humans tested by hierarchical variance as recorded by AMOVA.

| Level of Variance or Genetic Differentiation | Between Groups | Within Groups | Within Populations |
|---|---|---|---|
| **Vigilant *et al.* (1991) HVRI** | 0.76 | 42.24 | 57.0 |
| **Vigilant *et al.* (1991) HVRII** | 7.04 | 26.34 | 66.62 |
| **Single Species Hypothesis HVRI** | 21.01 | 33.82 | 45.18 |
| **Single Species Hypothesis HVRII** | 30.6 | 17.48 | 51.92 |
| **Subspecies Hypothesis HVRI** | 80.38 | 7.01 | 12.61 |
| **Subspecies Hypothesis HVRII** | 84.8 | 4.5 | 10.69 |

This result was further supported by a 100 % bootstrap score. In the case of the Vigilant *et al.* (h)(1991) HVRII phylogenetic analysis rooted with three chimp sequences from Morin *et al.* (2001) shown in Figure 11, the two neanderthal sequences yield the same result supported by a 100% bootstrap score.

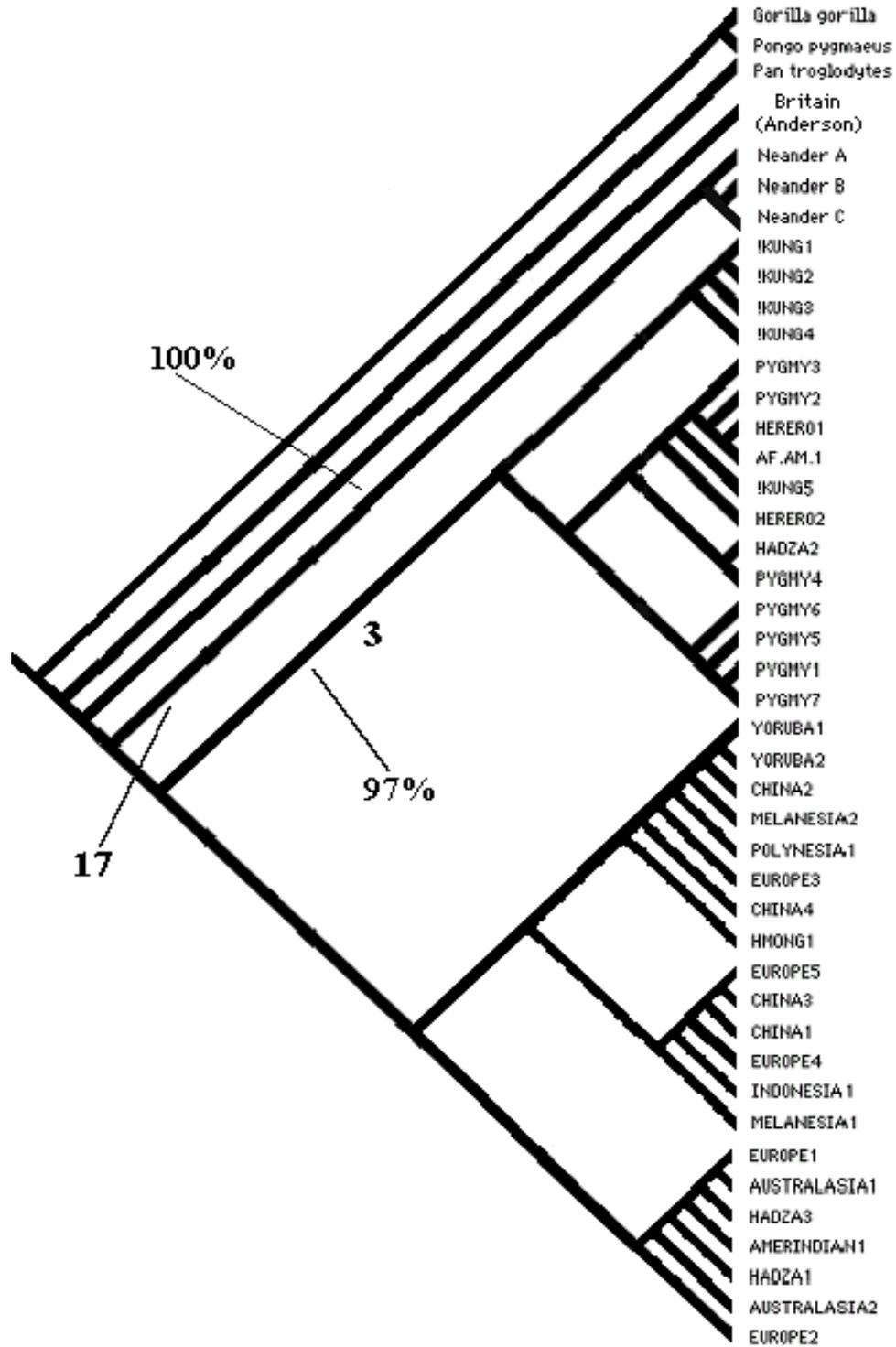

**Figure 10:** Results for the HVRI bootstrap analysis involving the Vigilant *et al.* (h)(1991) modern human and neanderthal sequences (bootstrap scores under 60% not shown).

Likewise, the 16S data from the Waters and Burridge (nh)(1999) single species dataset also yielded 100 % bootstrap support for the South American and Falkland Islands clade. The AMOVA percentages and tree statistics for a sample which combined neanderthal and modern humans was consistent with subspecies models using other great ape groups.

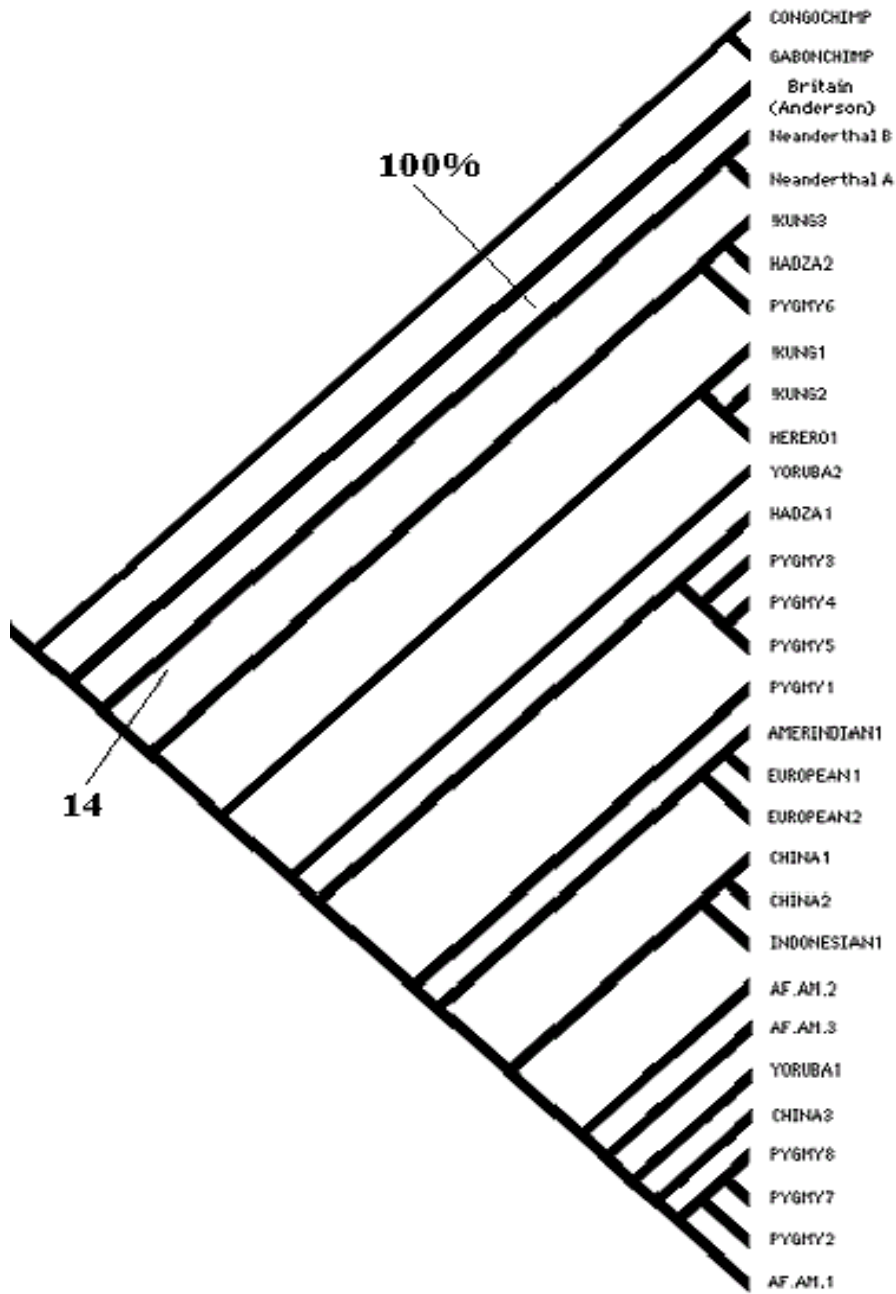

**Figure 11:** Results for the HVRII bootstrap analysis involving the Vigilant *et al.* (h)(1991) modern human and neanderthal sequences (bootstrap scores under 60% not shown).

This provides further support for the hypothesis that neanderthals represent a taxonomic group distinct from *H. sapiens* on either the subspecific level or greater. At the very least, these results show a paucity of evidence for supporting inclusion of neanderthal into modern *H. sapiens*. On the basis of what was found among specimens of *S. alpinus,* freshwater fishes in the southeastern United States, and the two experimental exercises, neanderthals should be classified as a subspecies which may have had the ability to interbreed with modern humans. Even though neanderthal is representative of a chronospecies, classifying it as a separate species based on these findings would conflict with the Biological Species Concept. Whether this interbreeding occurred throughout the entire geographic and temporal range of neanderthal is open to debate.

**Interpretive Analysis of Morphological Data**

The following section demonstrated how morphological characters range intraspecifically and how that relates to interspecific populations, while the concept of heterochrony was used specifically to investigate interspecific taxonomic discontinuity. In the case of *H. erectus*, a set of characters were used to determine whether the diversity found in this species represented either taxonomic discontinuity or intraspecific variation over space and time.

**Distinguishing Features Within *Homo erectus***

This section showed the results of a compiled table which examined morphological variance within the chronospecies *H. erectus*. Tests of continuity and the diagnostic nature of selected characters ultimately sought to establish how much variability should be allowed within a chronospecies without it being taxonomically significant.

To show the integrity of *H. erectus* as a single species and its relationship to *H. sapiens*, Table 14 was compiled for purposes of this thesis from data originally collected by Rightmire (1990). This included a systematic listing of eight *H. erectus* traits among 13 specimens. These data allowed for a comparison of trait size ranges between specimens within a single chronospecies and how they overlapped both over time and between regional populations. This information was intended for use as a generalized analogy to the size range measurements taken on the meristic characters of *Gymnotus* populations.

The ultimate purpose of this table was to compare variance in the range and mean of selected morphological characters from specimens of *H. erectus* over geologic time and portions of its geographic range. The amount of differentiation between taxonomic groups expressed for individual traits was determined by whether or the not the size range between specimens was found to overlap. Using this criterion, it was found that corpus height, corpus breadth, and torus thickness were variable over space and time, but not in any evolutionary meaningful way. By contrast, first molar breadth and occipital angle were not found to exhibit variation over space and time and appeared to be stable intraspecifically.

It was also found that corpus breadth was stable due to substantial overlap in the ranges of this character among several assemblages from a broad range of geographic and temporal locations. Meanwhile, cranial vault characters such as biauricular breadth and biasterionic breadth showed relatively wide variability in terms of their size range among different assemblages of *H. erectus*. Curiously, while molar size was stable within *H. erectus*, that same character has been found to be substantially reduced among *H. sapiens* (Rightmire 1993). This finding suggests that such a character has been plastic over evolutionary time but not necessarily intraspecifically (Brauer and Mbua 1992).

**Table 14:** Range of size (in mm) of selected traits in *Homo erectus* and how they change over space and time (compiled from Rightmire 1990).

| Assemblage | Age* | Torus Thickness | Biauricular Breadth | Biasterionic Breadth | Occipital Angle |
|---|---|---|---|---|---|
| Sangiran (Kabuh) | 600 | 14 to 19 | 124 to 139 | 119 to 126 | 100 to 102 |
| Ngandong | 53 | 12 to 16 | 128 to 146 | 124 to 126 | 94 to 104 |
| Zhoukoudian | 330 | 12 to 17 | 139 to 148 | 102 to 116 | ---------------- |
| Sangiran (Pucangan) | 105 | 12 | 124 to 132 | 120 to 124 | 104 |
| East Turkana | 1800 | 8 to 13 | 129 to 132 | 110 to 119 | 101 to 103 |
| Olduvai Bed 2 | 1400 | 19 | 135 | 123 | ---------------- |
| **Assemblage** | **Age*** | **Corpus Breadth** | **Corpus Height** | **First Molar Breadth** | **Second Molar Breadth** |
| Atlantic Morocco | 800 | 16 to 17 | 28 to 33 | 11.25 to 12.75 | 11.5 to 13 |
| Zhoukoudian | 330 | 15 to 18 | 28 to 35 | 10.5 to 12.75 | 11 to 13 |
| Olduvai Masek-Baringo | 500 | 20 | 30 to 33 | 11.25 to 12 | 11.5 |
| Ternifine | 530 | 16 to 20 | 32.5 to 37.5 | 12 to 12.75 | 12 to 13.5 |
| Olduvai Beds Three and Four | 770 | 20 to 22 | 28 to 35 | 11.5 to 12.5 | 11.5 |
| Sangiran (Pucangan) | 105 | 16 to 22 | 32.5 to 39 | 13 | 13.5 to 14.5 |
| East Turkana | 1800 | 19 to 20 | 28 to 33 | 10.75 to 12.75 | 11.5 to 13 |

*- average date (in thousands of years)

**Results of Neotropical Electric Fish Analyses**

To further capture the discrete nature of interspecific diversity, three morphometric measurements and two meristic counts were taken on various species groupings of Neotropical electric fish. The results of these two types of characters were considered separately among gymnotiform specimens due to differences in their diagnostic nature. For example, the ability of meristic counts to distinguish between taxonomic groups was based on comparing the distributions between species in histogram form. As with the *H. erectus* examples, differences among meristic characters included a lack of overlap in trait size. All diversity was compared graphically between groups to illustrate overall range distributions and growth trends.

**Anal-fin ray and pre-caudal vertebrae counts**

Specimens from the genus *Gymnotus* were examined by Albert and Miller (1995). The counts derived from this study were then graphically compared between species for purposes of this thesis. In addition, laboratory counts were made of anal-fin ray and pre-caudal vertebrae counts among several species in the genus *Gymnotus*.

Table 15 and Figure 12 contains the range counts and mean of anal-fin rays for eight species of *Gymnotus*. It was immediately apparent that the range of counts for *G. maculosus* was discontinuous with the ranges produced by specimens of *G. carapo* and *G. cataniapo*. This should be expected, since *G. maculosus* and *G. cataniapo* diverged approximately 50 million years ago (Albert, pers.comm.). The range for all other species tended to overlap, and thus did not produce significant taxonomic distance between the species groupings. It was also of note that all species of *Gymnotus* except for *Gymnotus cylindricus* displayed a continuous and unimodal distribution of anal-fin rays. This type of statistical distribution was used to determine the expected intraspecific among members of a given species for any meristic trait.

By contrast, *G. cylindricus* was found to exhibit a discontinuous distribution with a break at 195. While these type of breaks could be used to define a new species group within an established species, in this case they may also be due to sampling error.

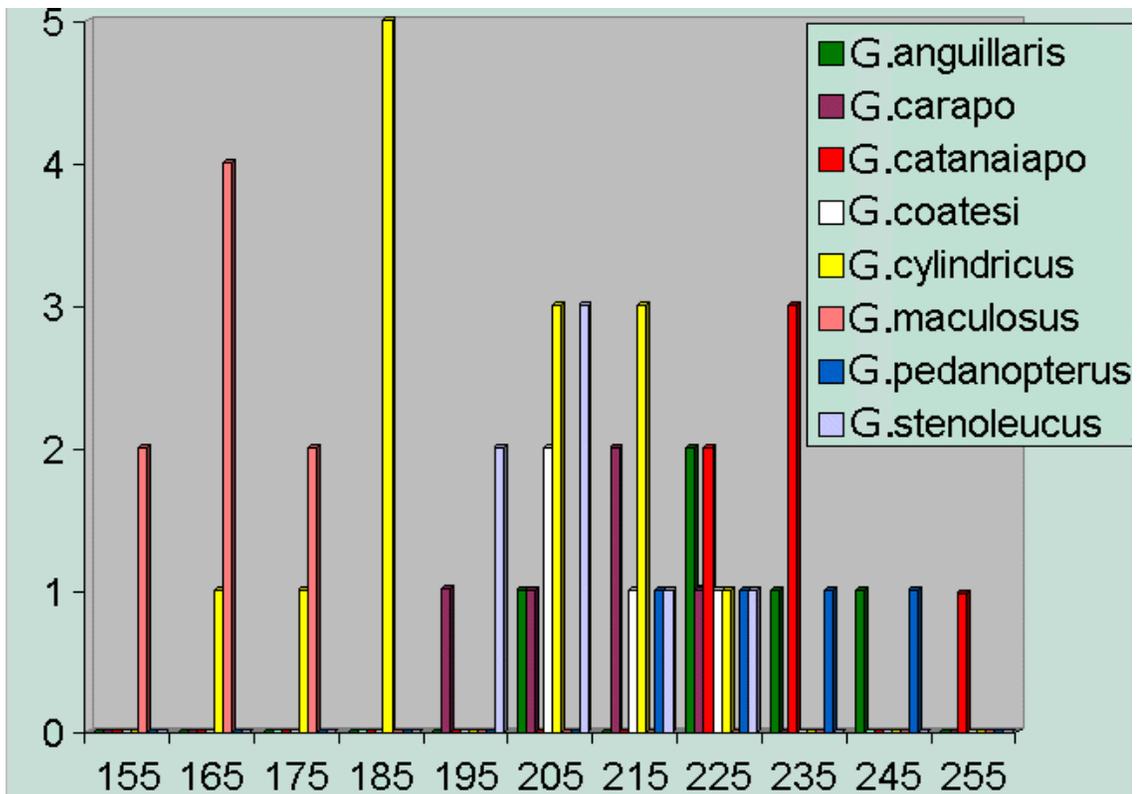

**Figure 12:** Range distribution in histogram form for diagnostic anal-fin ray counts for eight species of *Gymnotus* (Albert and Miller 1995).

Overall, these meristic characters were not only variable by species, but also exhibited a wide range of counts intraspecifically. However, it is of note that the species group *G. cylindricus* featured both the greatest number of specimens sampled and exhibited the widest trait size range found in this study. This may be interpreted as that although meristic characters have the ability to distinguish between species groups, they should be relied upon only in certain cases. For the most part, meristic traits appeared to grow and develop in a manner similar to morphological traits within *H. erectus*, but are not directly comparable.

This was further illustrated among the samples shown in Table 16 and Figure 13. In general, Figure 13 graphically demonstrated that the range of precaudal vertebrae counts have been found to differ in size between species groups of Gymnotus.

However, this was not always the case. For example, the range of counts for *G. coatesi*, *G. cataniapo*, and *G. stenoleucas* were all discontinuous. Meanwhile, the range of counts for *G. maculosus* and *G. cylindricus* were found to overlap.

**Table 15:** Anal-fin ray counts among eight species of *Gymnotus* (Albert and Miller 1995).

| Taxon | N | Mode |
|---|---|---|
| *G. anguillaris* | 5 | 225 |
| *G. carapo* | 5 | 215 |
| *G. catanaiapo* | 6 | 235 |
| *G. coatesi* | 4 | 205 |
| *G. cylindricus* | 14 | 185 |
| *G. maculosis* | 8 | 165 |
| *G. pedanopterus* | 7 | 235 |
| *G. stenoleucas* | 7 | 205 |

It was also shown that the analyzed anal-fin ray counts fell well within the generic range postulated by the Albert and Miller (1995) data. The laboratory counts shown in Figure 14 demonstrated that only *G. ayresi* and *G. varzea* exhibited discontinuous distributions relative to the rest of the taxa analyzed. However, the species groups *G. ayresi* and *G. varzea* both consisted of several specimens with 230 rays.

The case for *G. cylindricus* exhibiting a naturally discontinuous distribution of anal-fin rays with subdivision at 195 rays was not supported by the laboratory findings, as there were four specimens in the laboratory sample with 190 and one with 200 rays.

Once again, meristic characters provided only a general measure of distinguishing among taxa.

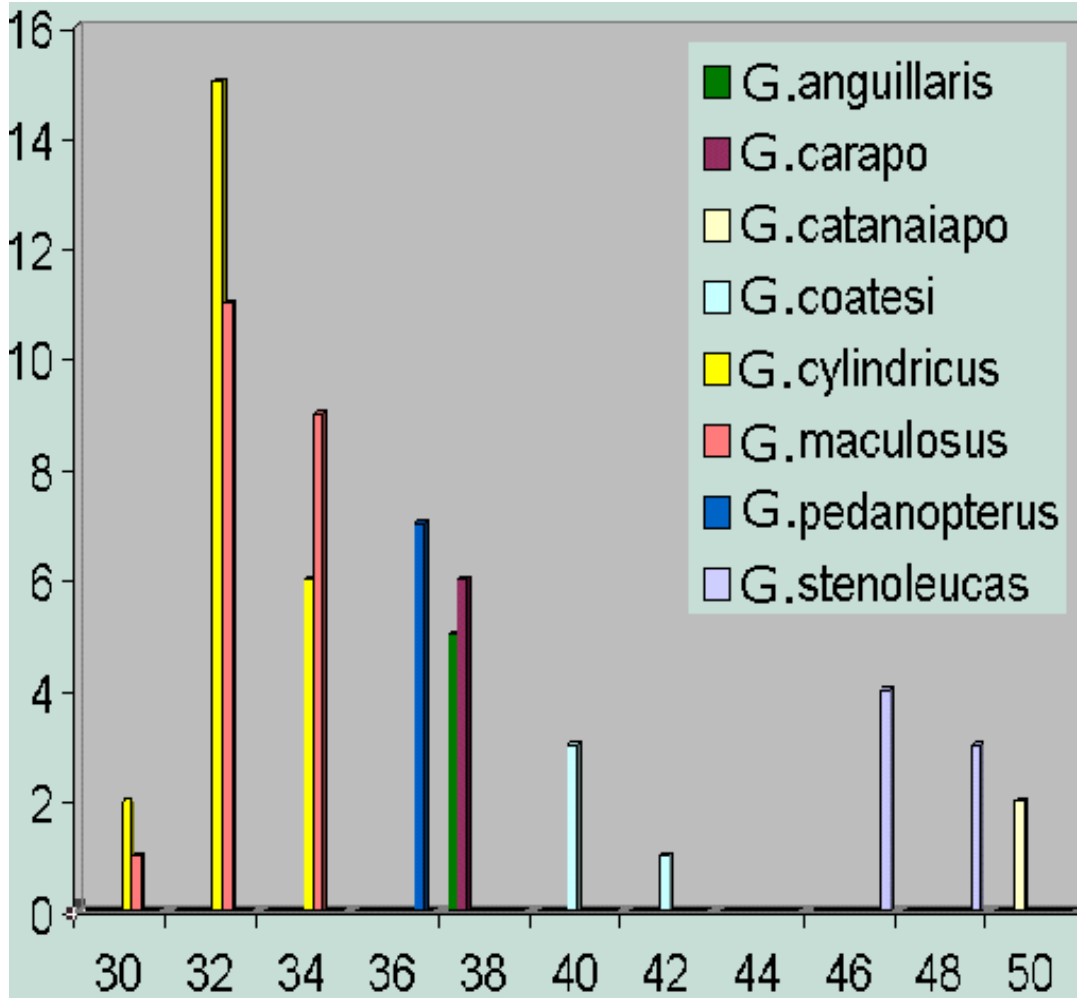

**Figure 13:** Range distribution in histogram form for distinctive precaudal vertebrae counts for eight species of *Gymnotus* (Albert and Miller 1995).

Among the anal-fin ray counts taken in the laboratory and shown in Figure 15, only the ranges for *G. varzea* and *G. ayresi* were found to be discontinuous with the other species analyzed. All other species sampled in Figure 15 exhibited overlapping distributions. In terms of pre-caudal vertebrae, the laboratory samples extended the range of counts in *G. maculosus* by two. In addition, the range of pre-caudal vertebrae counts for *G. varzea* were found to be distinct from all other species in the laboratory data.

**Table 16:** Precaudal vertebrae counts for eight species of *Gymnotus* (Albert and Miller 1995).

| Taxon | N | Mode |
|---|---|---|
| *G. anguillaris* | 5 | 38 |
| *G. carapo* | 6 | 38 |
| *G. catanaiapo* | 2 | 50 |
| *G. coatesi* | 4 | 40 |
| *G. cylindricus* | 23 | 32 |
| *G. maculosus* | 21 | 32 |
| *G. pedanopterus* | 7 | 36 |
| *G. stenoleucas* | 7 | 46 |

**Morphometric analysis of Neotropical electric fish diversity**

The *Sternopygus* species *xingu* and *aequilabiatus* were also differentiated by several key measurements of head and body proportions. For example, the mean body depth of *S. xingu* was 88.4 % of the head length (Albert and Fink 1996). This was greater than what was found among *S. aequilabiatus*, where the mean body depth was only 78.6 % of head length (Albert and Fink 1996). Other proportional standardized measurements also differed between the two species. For example, the ratio of mouth width to head length was found to be greater in *S. xingu* than in *S. aequilabiatus*, while the ratio of pectoral fin length to head length was less in *S. xingu* as compared to that of *S. aequilibiatus*. Meanwhile, the branchial opening to head length ratio was greater in *S.*

*xingu* than was found in *S. aequilabiatus* (Albert and Fink 1996). This difference was consistently large enough to propose these ratios as defining features of interspecific diversity.

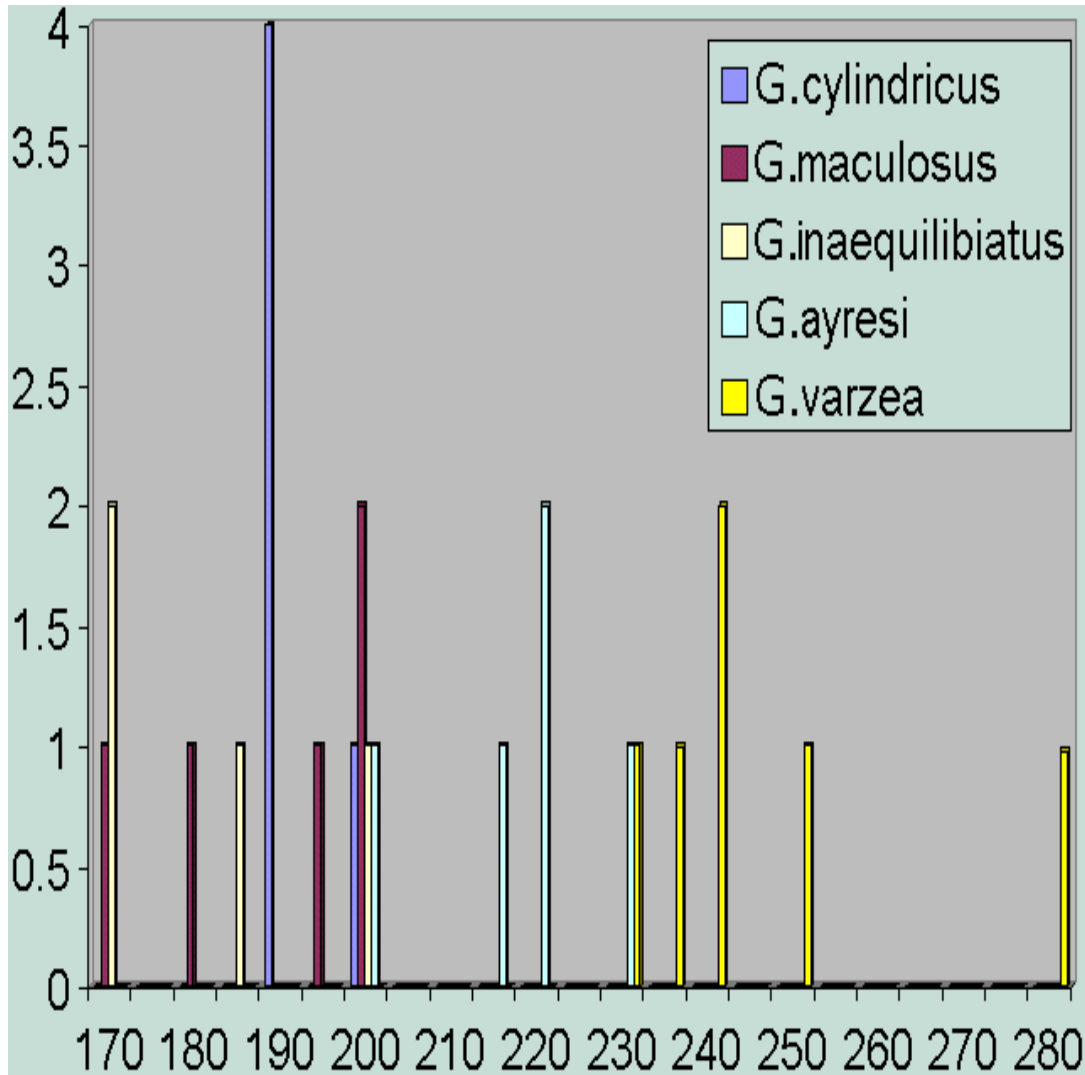

**Figure 14:** Anal-fin ray laboratory counts in histogram form among the same specimens of *Gymnotus*.

Similar trends were also apparent in the genus *Gymnotus*. On the graphs below, the growth trends for several species were compared for three different morphometric measurements. These growth trends were used to demonstrate both the process of heterochrony between species-level groups and the way in which growth processes can distinguish between taxonomically relevant populations.

**Table 17:** Anal-fin ray and pre-caudal vertebrae counts for recently analyzed specimens belonging to 5 species of *Gymnotus*.

| Taxon | N | Mode (anal-fin rays) | Mode (pre-caudal vertebrae) |
|---|---|---|---|
| *G. cylindricus* | 7 | 200 | 33 |
| *G. maculosus* | 5 | 200 | 34 |
| *G. inaequilibiatus* | 6 | 170 | 32 |
| *G. ayresi* | 5 | 220 | 35 |
| *G. varzea* | 8 | 240 | 38 |

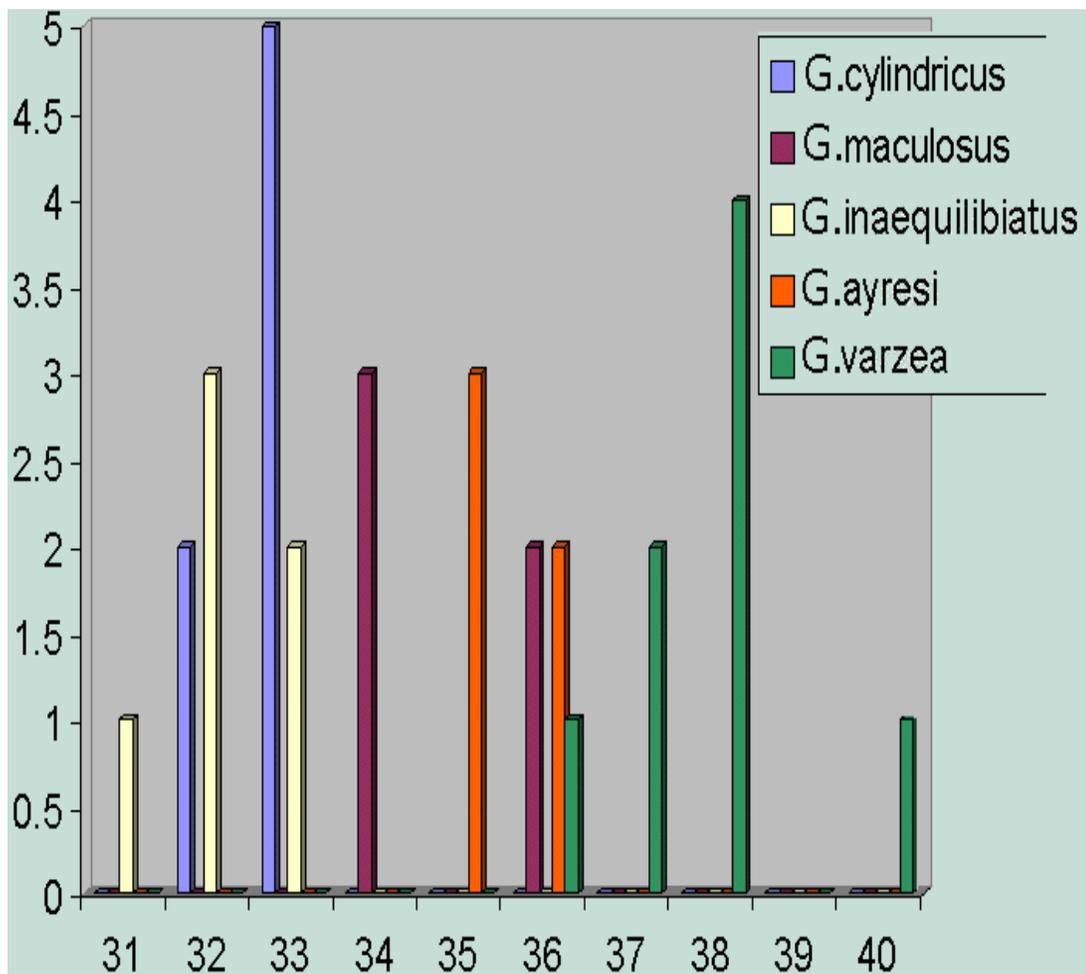

**Figure 15:** Pre-caudal vertebrae laboratory counts in histogram form among the same specimens of *Gymnotus*.

As was shown, each species formed a cluster that follows a trendline which was representative of a given sample's growth trajectory for that trait. Unlike humans, freshwater fishes exhibit indeterminate growth. This means that they grow throughout the course of their lives, but not during periods of ecological stress (Albert 1999b). Despite differences in the overall rate of growth for a specific character in a specific organism, by and large the traits for every species exhibit some degree of linear growth. Thus, using a linear regression technique has been found to provide a superior method for normalizing the resulting function.

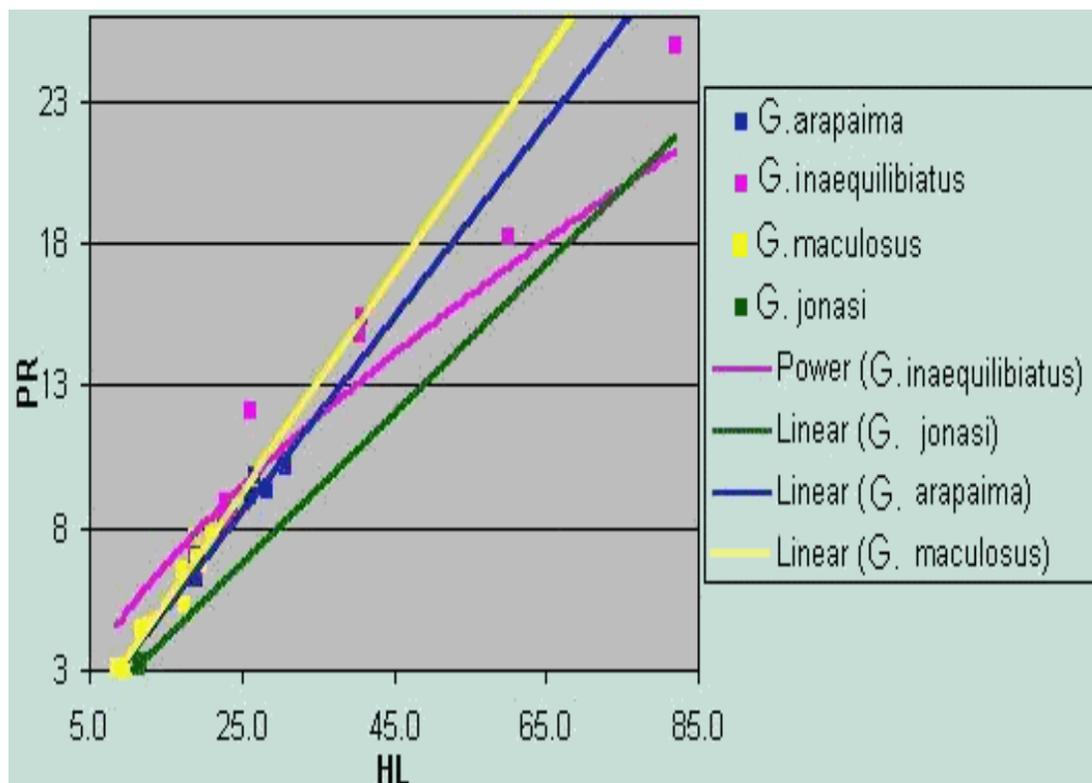

**Figure 16:** Bivariate plot of head length versus pre-orbital showing differential growth trends among selected species of *Gymnotus* (measurements in mm).

The only exception to this was when the distribution of variability in growth started in a linear fashion but grew more dispersed as the organism got larger. Such growth was characterized by the mouth width measurement of *G. inaequilibiatus* specimens. In such a case, the nature of growth was best characterized by a power function. This yielded slightly higher r-squared values than did a linear function for such distributions, and accounted for an increase of variance in the dependent variable as Head Length grew longer. In Table 18, the r-squared value and slope equation were listed for three traits and compared among four species. In Figures 16 through 18, the regression lines for each species were color coded on three graphs representing a single trait per graph. As was demonstrated by Figures 16 through 18, the growth trend for the pre-orbital region was the most divergent.

Overall, it was shown that the greatest amount of differentiation between species occurred when comparing pre-orbital measurements. Mouth width (MW) was also variable in a consistent manner, and so can also be considered an informative character for systematic analysis. However, mouth width measurements were as variable within species as it was interspecifically. Nevertheless, growth trends were derived from the analyzed specimens. Overall, the post-orbital region (PO) was found not to be informative as a representative for evolution between species. In general, this measurement has been the most evolutionarily conserved character observed thus far among Neotropical electric fish.

Not only were differences in the distribution of trait size observed between individual species, but the rate of growth for these characters over the lifespan of individual specimens also varied between species. The growth rates for head length, pre-orbital, and post-orbital measurements were compared by using a regression trendline as an indicator of growth over a specimen's lifespan. The resulting trend was then compared with trendlines derived from other species using the same methodology.

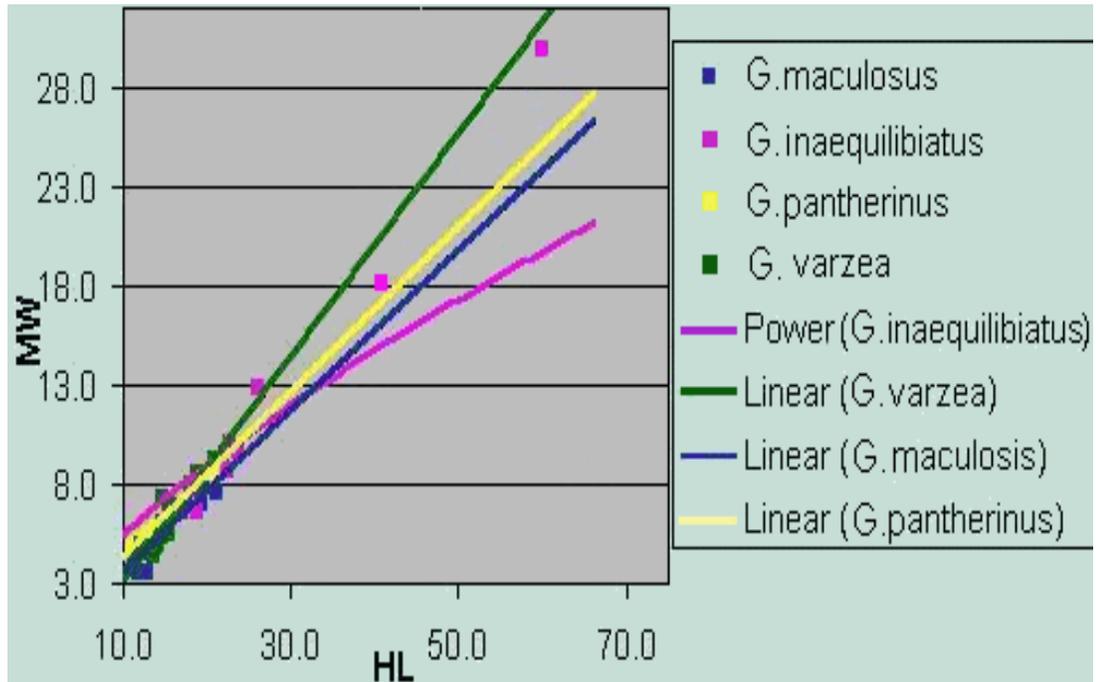

**Figure 17:** Bivariate clustering of head length versus mouth width showing differential growth trends among selected species of *Gymnotus* (measurements in mm).

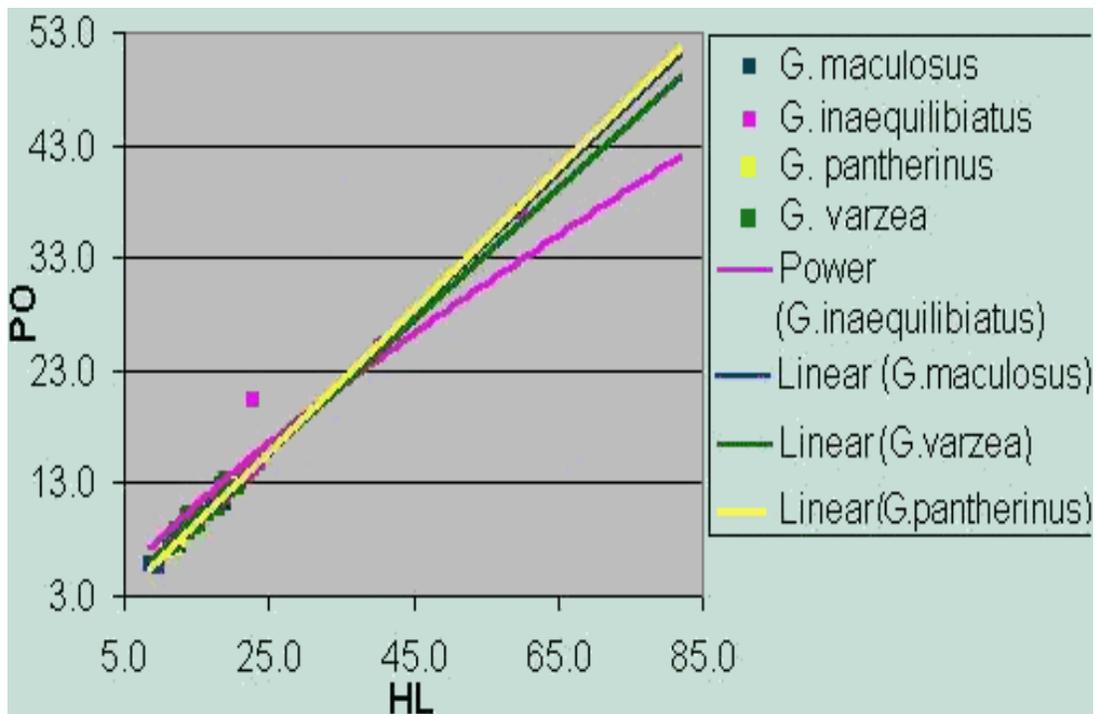

**Figure 18:** Bivariate clustering of head length versus post-orbital showing differential growth trends among selected species of *Gymnotus* (measurements in mm).

The growth trends for two of the most variable traits and one of the most stable evolutionary traits for selected species of *Gymnotus* were ultimately derived through qualitative but discrete states could be derived which neatly summarized the heterochronic relationships between taxa compared in Table 20. It was also found that for pre-orbital and mouth width, all taxa experienced an acceleration in growth ratewhen compared to *G. inaequilibiatus*. Thus, it appears from these comparisons that *G. inaequlibiatus* may be the ancestral species of all taxa compared in the analysis.

**Table 18:** Heterochrony states for selected *Gymnotus* species derived from graphed data.

| Taxon | Pre-Orbital | Post Orbital | Mouth Width |
|---|---|---|---|
| *G. varzea* | NA | Little difference | Starts later, grows much faster |
| *G. jonasi* | Starts later, grows faster | NA | NA |
| *G. maculosus* | Starts later, grows much faster | Little difference | Starts later, grows faster |
| *G. pantherinus* | NA | Little difference | Starts later, grows faster |
| *G. arapaima* | Starts later, grows faster | NA | NA |

Using *G. inaequilibiatus* as the outgroup taxon, the taxa featured in Table 18 were analyzed in a phylogenetic analysis to show the evolutionary relationships between taxa as they were affected by changes during ontogeny. When a character exhibited "little change", it was considered to be ancestral.

As was shown in Figure 19, there was a resolvable amount of difference between the taxa. This was indeed found to be the case even though the phylogeny included only three characters. Finally, the slopes for each trait derived using a regression analysis were also compared to yield a more quantitative description of the same phenomena. As should be expected, the "plastic" traits showed differences in the overall growth trend, while the evolutionarily conserved traits exhibited little difference relative to one another. Overall, it has been shown that evolutionarily relevant changes were expressed by differential growth trends that lead from ontogeny to adulthood. This was especially true of the pre-orbital region, which was found to be derived in three of the five ingroup taxa.

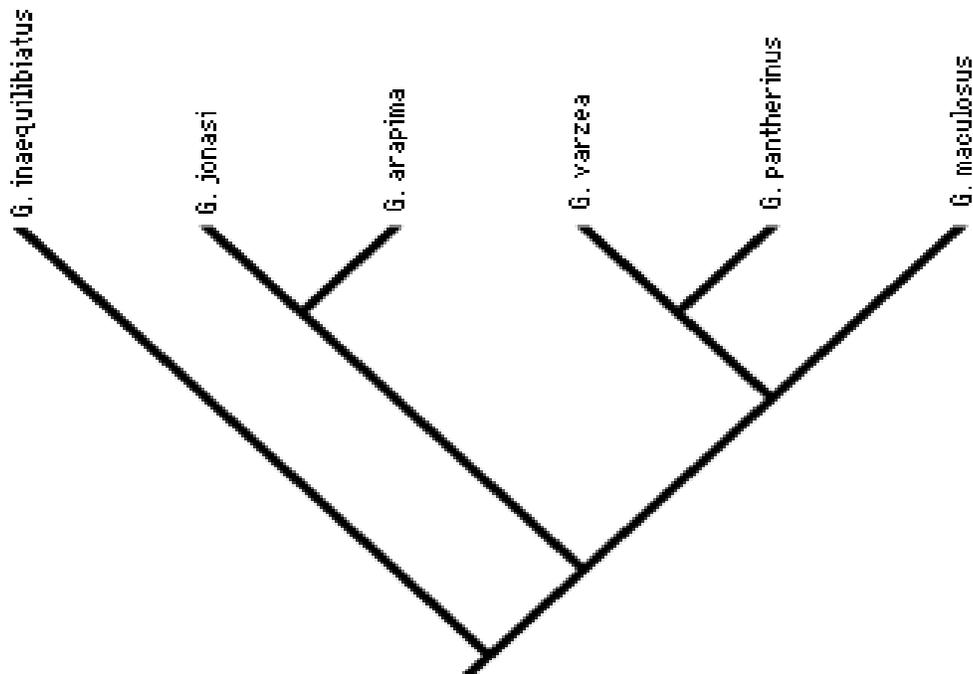

**Figure 19:** Phylogenetic analysis of differentiation due to changes during ontogeny.

In conclusion, it was found that intraspecific molecular variation can either be panmictic or discontinuous. When an intraspecific sample was found to be discontinuous, that group was classified as exhibiting taxonomically relevant discontinuity. In this thesis, modern *H. sapiens* were found to be panmictic while samples incorporating both modern human and neanderthal samples were discontinuous. This was in comparison to two discontinuous and one panmictic fish species.

**Table 19:** Heterochrony states for selected *Gymnotus* species.

| PR vs HL* | Power Slope | Linear Slope | Correlation Coefficient |
|---|---|---|---|
| *G. arapaima* | | y= 0.34x - 0.08 | R^2= 0.91 |
| *G. inaequilibiatus* | y= 1.05x^ 0.68 | y= 0.21x + 4.67 | R^2= 0.61 |
| *G. maculosus* | | y= 0.39x - 0.40 | R^2= 0.92 |
| *G. jonasi* | | y= 0.26x + 0.11 | R^2= 0.96 |
| **MW vs HL*** | | | |
| *G. maculosus* | | y= 0.40x - 0.41 | R^2= 0.89 |
| *G. inaequilibiatus* | y= 1.04x^ 0.72 | y= 0.26x + 4.87 | R^2= 0.43 |
| *G. pantherinus* | | y= 0.42x + 0.05 | R^2= 0.90 |
| *G. varzea* | | y= 0.56x - 2.47 | R^2= 0.87 |
| **PO vs HL*** | | | |
| *G. inaequilibiatus* | y= 1.32x^ 0.79 | y= 0.49x + 5.17 | R^2= 0.61 |
| *G. varzea* | | y= 0.58x + 1.02 | R^2= 0.85 |
| *G. maculosus* | | y= 0.62x + 0.06 | R^2= 0.98 |
| *G. pantherinus* | | y= 0.63x + 0.03 | R^2= 0.99 |

*- PR= Pre-orbital Length, MW= Mouth Width, HL= Head Length, and PO= Post Orbital

It was also found that meristic traits are distinguishing characteristics of interspecific diversity only when they yield discontinuous distributions between species groups. In addition, the growth trends for evolutionarily significant morphological traits change in onset and rate during the juvenile phase. This morphological change eventually amounted to a recognizable degree of difference between interspecific populations.

# DISCUSSION AND CONCLUSIONS

This section provided several examples of how findings from the previously analyzed data relate to other studies and theories. Fish species from the southeastern United States were presented to illustrate intraspecific variation and discontinuity using phylogeographic methods. A short discussion on implications of the neanderthal experimental exercises also contributed to an understanding of molecular interspecific variation. Finally, applications of heterochrony to human evolution were discussed, in particular the interspecific populations which result from processes initiated during ontogeny. This was used in relating morphological changes to human evolution.

**Intraspecific Discontinuity in Human and Fish Populations**

Given the aforementioned requirements to which morphological data must conform, the reanalysis of human craniometric data from Howells (1973) yielded varying degrees of evidence for intraspecific subdivision. While Howells (1973) was able to produce evidence for geographically specific groupings using a full suite of morphometric traits, they could not be clearly defined using the traits that characterize prognathism. Overall, there must be a reasonably significant amount of gene flow between human subpopulations in order to maintain the specific integrity of *H. sapiens*. Rogers (1984) has even postulated that modern humans are a group of variants loosely connected by a tendency for adapting to their environment rather than a genetic unity. If this model is accurate, extreme variations in polymorphism can be expressed without the inevitability of intraspecific discontinuities. In light of this, intraspecific subdivision must be identified with either explicit reference to geographic location or habitat differentiation. The geographic definition was found to be strongly supported by the phylogeography of *S. alpinus*.

**Implications of Molecular Diversity**

The following section discusses the results of several studies which further illustrate the types of diversity exhibited by the eight analyzed datasets. The nature of molecular discontinuity based on geographic distribution was discussed by introducing several cases involving freshwater fishes from the southeastern United States. Ultimately, these studies further supported the hypothesis that looking at subdivided fish species provides a heuristic tool for detecting patterns of molecular diversity within *Homo sapiens*.

**Nature of Discontinuous Intraspecific Subdivision**

In this study, it was found that all *H. sapiens* populations analyzed consistently yielded a discontinuous population distribution with more variance than within populations. These findings can be contrasted with what was found to exist among several Sunfish and Bowfin populations. These species reside in river basins that range across the southeastern United States. Bermingham and Avise (1986) sampled 75 Bowfins (*Amia calva*) from across the southeastern United States, and found that 13 distinct haplotypes extended from South Carolina to Louisiana. These analyses yielded two major assemblages of *Amia calva* that were both genetically and geographically distinct. Separated by the Appalachicola river basin, an eastern group of nine haplotypes emerged, along with a western haplogroup consisting of four haplotypes. The geographic discontinuity was important because at least four restriction site changes were found to distinguish any eastern haplotype from a western one. In particular, a single haplotype was found to be geographically widespread in the east, present in 30 of 59 specimens. Bermingham and Avise (1986) also showed that demes of *Lepomis microlophus* (Redear Sunfish) exhibit spatial discontinuity between eastern and western haplotypes with a mutational distance of 17 steps. This was greater than the distance encountered when exchanging monophyletic groupings consisting of neanderthal and modern human samples. This type of deep regional structuring was also found among discontinuous populations of *G. maculatus*, but stands in contrast to what was found within the Vigilant *et al.* (1991) modern human sample.

For any species that has expanded dramatically from single refugia or place of origin in recent evolutionary times, internal phylogeographic differentiation should be quite limited. Johnson (1983) assumed that humans are better characterized as exhibiting a partial spatial separation model of phylogenetic continuity. While there has been some isolation between local human populations, that isolation has not resulted in significant discontinuities. While a lack of intraspecific subdivision has been found within *H. sapiens*, more specific and temporary examples of intraspecific discontinuity among modern humans may be produced by the interaction of linguistic and cultural factors.

**Cultural Factors and Geographic Variation**

Spatial and taxonomic discontinuity between populations of modern humans can primarily be related to cultural and linguistic boundaries. These constraints differentiate modern humans from fish by resulting in noticeably distinct but temporary and shifting clines within an otherwise panmictic species range (Barbujani and Sokal 1990). Taking this into consideration, Long *et al.* (1990) proposed a measure to compare the patterns of genetic variation with demographic, geographic, and linguistic variation among linguistically-differentiated populations in Papua New Guinea. In their model, genetic distances were compared to the level of endemicity of the populations for each geographic subdivision sampled. Endemicity was defined as the probability that both parents were born in the same subdivision. Following this assumption, parishes with a low endemicity were found to be less divergent than parishes associated with high endemicity.

Long *et al.* (1990) found that Kalam-speaking regions were far less internally differentiated than Gainj-speaking regions. Slatkin (1987) observed that within all animal populations, subpopulations either internally recombine or remain stable over time. Cases where subdivisions occurred on the basis of ethnicity or clan membership usually involved groups that fissioned every few generations. This example also shows that cultural and linguistic factors affected the distribution of intraspecific variation, but not to an extent which was taxonomically significant.

**Discussion of Neanderthal Taxonomic Status**

The cladistic exchange exercise yielded evidence for limited interbreeding between a distinct subspecies of neanderthal and modern *H. sapiens.* Likewise, the taxonomic variance experimental exercise yielded higher between group than within population variance when neanderthal was treated as a group distinct from all modern human populations. The nature of these variance components provided evidence for neanderthal constituting a subspecies separate from modern humans.

A smaller-scale test looking at the exact number of mutational steps between haplotypes may provide a better estimate of the taxonomic relationship between neanderthal and *H. sapiens*. Comparing the Anderson *et al.* (1982) reference sequence with four modern human sequences from Vigilant *et al.* (1991) and the three HVRII neanderthal sequences clearly demonstrated the divergent nature of neanderthal. For example, the distance between the Anderson *et al.* (1982) reference sequence and tested European sequences ranged between one and three mutations. When a Chinese sequence was compared to the reference sequence, there was a difference of seven mutational steps. A similar comparison with one !Kung sequence yielded a difference of 10 steps.

According to Krings *et al.* (1997) and the results of this comparison, the distance between the reference sequence and the neanderthal sequence from Germany was found to be 27 mutational steps. These findings imply a gap between the full range of modern human diversity and neanderthal diversity of roughly 17 steps. Furthermore, the !Kung and neanderthal sequences were found to share some polymorphisms which were not present in other human sequences. Populations of *Opsanus tau* (Marine Oyster Toadfish) and *Geomys pinetis* (Pocket Gopher) have demonstrated that this type of mitochondrial discontinuity could be due to either geographic isolation or the extinction of intermediates (Templeton 1989). This may also imply that both the !Kung and neanderthal are representative of ancient diversity.

Nordborg (1998) used a probabilistic method concerning whether or not neanderthals constitute a portion of the modern human panmictic distribution. Since interbreeding between two taxonomically distinct groups is an evolutionarily primitive characteristic, Mismatch distributions measuring ancient levels of gene flow and coalescent methods were utilized instead of a phylogenetic approach. It must be kept in mind that all estimates of gene flow based on coalescent methods are contingent upon a large sample size. Because of this, Nordborg (1998) concluded that the chances of correctly detecting interbreeding that took place before the time in which the effective modern human population size began to grow exponentially is close to zero. Using mitochondrial polymorphisms, Nordborg (1998) estimated the most recent common ancestor between the modern human and neanderthal lineages to have existed between 800,000 and 150,000 years ago depending on the rate of population growth used in the simulation. Estimates of these coalescence times presumed a perfect knowledge of mitochondrial genealogy, for there is considerable uncertainty involved with estimating this history. This was based on the observation that even though human mitochondrial DNA genealogies are star-shaped, the analysis of multiple nuclear loci fail to yield a similar pattern (Harding 1997). While both Nordborg's (1998) models and the experimental exercises presented in this thesis resulted in various amounts of support for interbreeding, a strictly phylogenetic approach for identifying the taxonomic limits of both modern humans and neanderthals may be most effective.

Nordborg (1998) also worked from the more basic assumption that neanderthal mitochondria are several times more ancient than that of modern humans. This finding has lead to the assumption by people such as Ward and Stringer (1997) that modern humans totally replaced neanderthals without the occurrence of interbreeding. Conversely, Enflo *et al.* (2001) suggested that reproductive differences between neanderthal and modern humans between different regions could mask any inbreeding between neanderthal and modern humans. For purposes of running a Monte Carlo simulation, Nordborg (1998) assumed that the average neanderthal mother was at a reproductive disadvantage and had less than two surviving children. If modern human females were assumed to be producing more than two children on average per generation, neanderthal populations on the whole would have been at a reproductive disadvantage to modern human populations migrating into a given region. As the center of both neanderthal populations and climatic extremes during the Pleistocene, Europe would have been especially affected by these reproductive differences and migrations. Thus, neanderthal genetic contributions to the modern human gene pool would be invisible to the modern observer.

Conclusions reached from the findings of Krings *et al.* (1997) and Enflo *et al.* (2001) were also tested by Nordborg (1998) to determine whether or not support for the replacement hypothesis is simply the result of a Type I statistical error. Ultimately, Nordborg (1998) found that the replacement model exhibited a higher likelihood of occurrence on average than models assuming interbreeding. It was found that if neanderthals merged with the modern human lineage and constituted 25 % of the resulting population, the probability that neanderthal haplotypes were lost through drift was 52%. To arrive at these statistics, Nordborg (1998) presumed that neanderthals existed from the previously estimated coalescence times with the original human lineage until modern humans were encountered approximately 68,000 years ago. These two populations would then have interbred to form a randomly mating single species.

Finally, examples of great ape diversity and the African elephant case study of Roca *et al.* (2001) can be used to illustrate the parameters of modern *H. sapiens* diversity. Such studies can be used comparatively in much the same way fish species were used to establish the parameters of intraspecific diversity. The hypothesis that two African elephant species can be derived from the current species *Loxodonta africana* was found to be supported by a high genetic differentiation (F*st*) value (Roca *et al.* 2001). An extremely high value of .95 was combined with evidence that the hypothesized members of each proposed species inhabited separate phylogenetic clades supported by bootstrap scores of 95 %.

Likewise, Zhang *et al.* (2001), analyzed two geographically distinct subspecies of *Pongo pygmaeus*. Orangutans from Borneo are called *P. pygmaeus pygmaeus*, while Sumatran orangs are called *P. pygmaeus abelii*. This taxonomic differentiation was traditionally based upon unique morphological, behavioral, and cytogenetic characteristics. There has also been little evidence of a recent genetic bottleneck among either of the subspecies, as their effective population sizes has been found to be on the order of 10,000 individuals. When both of these subspecies were combined into a single 16S mitochondrial DNA sequence dataset and run through Arlequin (Schneider *et al.* 2000), the F*st* value was found to be .734. This was comparable to a F*st* of .885 yielded when a combined sample of chimpanzee (*Pan paniscus*) and Bonobo (*Pan troglodytes*) was analyzed by Horai *et al.* (2001). All of these populations are comparable to the taxonomic relationship between neanderthal and modern humans proposed in this thesis, and yield comparable genetic differentiation and bootstrap statistics[1].

**Character Differentiation and Applications to Human Evolution**

This final section focuses on comparisons between what was found among analyses of Neotropical electric fish with major issues and contentions in human evolution. The nature of meristic characters and interspecific comparisons were discussed first, which was followed by the implications of heterochrony in human evolution.

Meristic characters were only found to be marginally useful in distinguishing between taxonomic groups. While in some cases the distribution of traits for individual species groups were discontinuous, cases where the ranges of all taxa sampled overlapped were more common. In this case, a given trait could still be used as an evolutionarily useful character providing counts could be obtained from an ancestral taxon. Likewise, growth trends for evolutionarily plastic characters such as the pre-orbital region and mouth width of several *Gymnotus* species were shown to be divergent. This implies that changes in form among these organisms occurs during ontogeny.

Heterochrony as a mechanism for expressing evolutionary change between interspecific groups is apparent in human evolution as well. Ontogenetic variables within the genus *Homo* have traditionally included gestation length, dental eruption timing, reproductive maturation, and life span (Leigh 2001). As is the case with investigating heterochrony among gymnotiform groups, the observation of significant taxonomic discontinity involves regular increments of change in these variables. Such traits are distinctive in the same way as other modern human apomorphies (Leigh 2001). For example, alveolar prognathism and cranial vault shape have been variable throughout the course of human evolution. Differential rates of growth in these traits among reproductively isolated populations has resulted in interspecific diversity over time.

In addition, extention of the preadult period initiated by insertions of novel growth periods into the life cycle has affected modern human behavior in ways such as increased parental investment (Leigh 2001). This development of a long preadult period is the most obvious example of heterochrony in human populations (Shea 1989), and is what differentiates human applications of heterochrony from applications of this theory to Neotropical electric fish specimens.

Finally, interspecific relationships based on heterochronic processes have been exemplified in the hominid braincase. Falk (1987) found that the lunate sulcus, located between the frontal and parietal regions of the skull, were enlarged and changed position among *Homo* specimens. The physiological function of this character is to keep the brain thermoregulated. Thus, the lunate sulcus is hypothesized to be characteristic of general trends in accelerated brain growth.

Enlarged lunate sulci are not simply definitive of a entire genus, but especially characterizes growth in the Broca's area. Change in the size of this region often implies the development of speech capabilities (Wilkins and Wakefield 1995). Similarly, a form of heterochrony called time hypermorphosis has been found to affect these same hominid populations in Africa 2.5 million years ago (Vrba 2001). This resulted in derived populations which exhibited morphological features both larger and smaller than their ancestral forms. Ultimately, this was found to drive hominid encephalization.

**Conclusions**

A comparative approach which relied upon taxonomy to interpret subdivisions found within populations of humans and fish yielded several general results. Through the observation of human and fish data, *H. sapiens* was found to exhibit little significant intraspecific discontinuity and larger within population variance. By contrast, two of the three fish species exhibited greater between group variance and discontinuity supported by high bootstrap values. Upon performing two experimental exercises on interspecific molecular data, it was found that neanderthals and modern humans exhibited not only discontinuity when combined into a single species sample, but also taxonomically relevant differences when samples from each group were compared with one another.

Even though heuristic racial groupings have been found to exist within *H. sapiens*, it must be made clear whether this is based on systematic subdivision or the mere existence of polymorphism. In fact, it may be that analyzing Y-chromosome or microsatellite data is more effective than using mitochondrial loci for resolving intraspecific subdivision. Given this concern, it should be noted that geographically based intraspecific groups can be resolved through means other than analyses of mitochondrial DNA. While Nei and Roychoudhury (1993) found evidence for intraspecific modern human discontinuity by sampling 29 shared polymorphic loci from the nuclear genome among populations from five continents, the average amount of admixture allowed for all the populations used in this study was under 10% of the total population.

The analysis of meristic characters and the application of heterochrony to morphometric characters showed that traits with discontinuous numeric distributions and differential growth trends are most characteristic of morphological interspecific subdivision. The schema of Burt (2001) is important for illustrating the nature of this finding. This model proposes that both within a chronospecies and between species groups, traits can exhibit either evolutionary stasis or thrust. Traits which exhibit stasis intraspecifically and thrust interspecifically should be most useful in determining taxonomic discontinuity.

Ultimately, there were several assumptions inherent in recognizing intraspecific and interspecific discontinuity within *H. sapiens* by applying observations made among taxonomically subdivided fish populations. Nevertheless, this thesis was found to be a successful exercise in investigating the parameters of species-level diversity within the genus *Homo* in an objective and unique manner.

**Footnotes:**
[1]- It is of note that percentages of variation and the high genetic differentiation value (F*st*) of .87 associated with treating neanderthal as a subspecies is strong evidence for the existence of taxonomic subdivision between these groups.